\documentclass[aps,twocolumn,floatfix,superscriptaddress,showpacs,showkeys]{revtex4-1}
\usepackage{epsf,amsmath,amssymb,verbatim,color,multirow,pifont,mathrsfs,soul,times}
\usepackage{graphicx}
\usepackage[linktoc=none,colorlinks=true]{hyperref}

\begin{document}

\title{Stable Chimeras and Independently Synchronizable Clusters}

\author{Young Sul Cho}
\affiliation{Department of Physics and Astronomy, Northwestern University, Evanston, IL 60208, USA}
\affiliation{Department of Physics, Chonbuk National University, Jeonju 561-756, Korea}

\author{Takashi Nishikawa}
\affiliation{Department of Physics and Astronomy, Northwestern University, Evanston, IL 60208, USA}
\affiliation{Northwestern Institute on Complex Systems, Northwestern University, Evanston, IL 60208, USA}

\author{Adilson E. Motter}
\affiliation{Department of Physics and Astronomy, Northwestern University, Evanston, IL 60208, USA}
\affiliation{Northwestern Institute on Complex Systems, Northwestern University, Evanston, IL 60208, USA}

\begin{abstract}
Cluster synchronization is a phenomenon in which a network self-organizes into a pattern of synchronized sets. It has been shown that diverse patterns of stable cluster synchronization can be captured by symmetries of the network.  
Here we establish a theoretical basis to divide an arbitrary pattern of symmetry clusters
into {\it independently synchronizable cluster sets}, in which the synchronization stability of the individual clusters in each set is decoupled from that in all the other sets.
Using this framework, we suggest a new approach to find permanently stable chimera states by capturing two or more symmetry clusters---at least one stable and one unstable---that compose the entire fully symmetric network.
\end{abstract}

\onecolumngrid\hfill 
{\small {\it Phys. Rev. Lett.} {\bf 119}, 084101 (2017)

\hfill\href{https://doi.org/10.1103/PhysRevLett.119.084101}{https://doi.org/10.1103/PhysRevLett.119.084101}\hspace{-1mm}}
\bigskip\twocolumngrid

\maketitle

Synchronization is a collective network behavior in which the states of the interacting units evolve in step with each other~\cite{sync_book}, as observed in animal flocking~\cite{PRL_vicsek}, the coordinated firing of neurons~\cite{Engel2001,Varela2001,Axmacher2006}, and the synchronous operation of power generators~\cite{Nphys_motter}.
Beyond the complete synchronization of all units, significant progress has been made on understanding more complex forms of synchronization, including cluster synchronization~\cite{Chaos_syncbasin,pecora_ncomm2014,pecora_arxiv2015,EEP,Sorrentino2007,Dahms2012,Chaos_cluster} and chimera states~\cite{Kuramoto_chimera,PRL_Abraham,PNAS_chimera,Abrams_chimera,Abrams2015,Wolfrum2011kd,Suda2015,Panaggio2016,Hart2016,Hagerstrom2012,Tinsley2012,Bick2016,Bohm2015,star_chimera}.
In particular, cluster synchronization (CS), in which clusters of nodes exhibit synchronized dynamics, has seen recent breakthroughs:
rigorous relations based on group theory have been established between patterns of synchronous clusters and the symmetries of the network structure~\cite{pecora_ncomm2014, pecora_arxiv2015}.
Network symmetry can be used to explain various forms of collective behavior, such as remote synchronization, in which two nodes are synchronized despite being connected only through asynchronous ones~\cite{PRL_remote}, and isolated desynchronization, in which the synchronization of some clusters is broken without disturbing other clusters~\cite{pecora_ncomm2014, pecora_arxiv2015, PRE_partial1, PRE_partial2}.

Chimera states, which are characterized by the coexistence of both coherent and incoherent dynamics within the same state, are also intimately related to symmetry.
Since the initial discovery~\cite{Kuramoto_chimera} and subsequent analysis~\cite{Abrams_chimera} of such states, numerous studies have found---numerically, analytically, and experimentally---that chimera states can be observed in a wide range of systems (see the review in Ref.~\cite{Abrams2015} and the references therein).
However, it was recently found that chimeras in finite-size networks can be long-lived but transient states~\cite{Wolfrum2011kd}
(i.e., the system will eventually settle onto a simpler state, 
such as complete synchronization).
This raised a fundamental question: are permanently stable chimeras possible with a finite number of oscillators~\cite{note2}?
Evidence for the affirmative answer has so far been limited to numerical simulations~\cite{Suda2015,Panaggio2016,Bohm2015} (notable exceptions are two case studies with stability analysis: one for ``weak'' chimeras in bistable populations of phase oscillators~\cite{Bick2016}
and the other for a four-node network of delay-coupled opto-electronic oscillators~\cite{Hart2016}).
Our approach for addressing this problem is to identify symmetry-based ``templates'' for chimeras: a partition of the network into synchronization clusters including both a stable one and an unstable one.

\begin{figure}[b]
\includegraphics[width=0.9\linewidth]{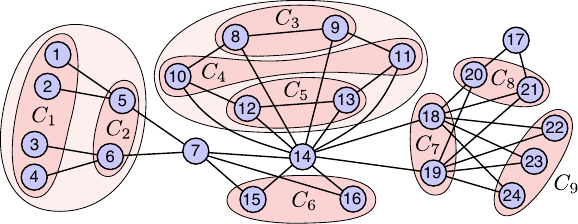}
\vspace{-5pt}
\caption{
Grouping of 
symmetry clusters in a CS pattern for a $24$-node network 
(detailed in Supplemental Material~\cite{SM}, Sec.~I).
}\label{Fig:Independent_Intertwined_examples_schematic}
\end{figure}

In this Letter, we develop a general framework that can be used to systematically search for such partitions and, moreover, to characterize any symmetry-based CS patterns in a network.
Specifically, we establish that, for any given partition of a network into symmetry clusters, we can uniquely identify groups of clusters in which those in the same group must have the same stability while those from different groups can have different stability (see Fig.~\ref{Fig:Independent_Intertwined_examples_schematic} for an example).
In particular, a single cluster forming a group by itself would be a candidate for the stable cluster in a permanently stable chimera state.
We show that these groups can be computed efficiently and provide examples of finding permanent chimeras using this approach.
The decoupling of stability between different groups of clusters is derived using a cluster-based coordinate transformation, which is much simpler and is demonstrated to be faster to compute than the one based on the 
group-theoretical characterization
of the network's symmetries~\cite{pecora_ncomm2014, pecora_arxiv2015}.

We consider networks of $N$ nodes, each representing an $n$-dimensional oscillator $\bold{x}_i$ governed by
\begin{equation}
\dot{\bold{x}}_i(t)=\bold{F}(\bold{x}_i(t))+\sigma\sum_{j=1}^N A_{ij}\bold{H}(\bold{x}_j(t)),
\label{pecora_rate}
\end{equation}
where $\bold{F}(\bold{x})$ determines the uncoupled dynamics of the nodes, $\sigma$ is the global coupling strength, $A = (A_{ij})_{1\le i,j\le N}$ is the adjacency matrix of the network, and $\bold{H}(\bold{x})$ is the coupling function. 
For concreteness we consider undirected and unweighted networks, but our theory can be extended naturally to directed and weighted networks~\cite{note1}.
A {\it CS pattern} for this system is a partition of the network into clusters of oscillators that are in stably synchronized states characterized by $\bold{x}_i(t)=\bold{x}_j(t)$ for all $i$ and $j$ in the same cluster.
The network can be multi-stable and thus allow for multiple CS patterns.
For a given network structure, the specific synchronization pattern realized is determined by the initial condition defined by $\bold{x}_i(0)$ for all $i$.
A large set of candidate CS patterns---whose stability depends on $\bold{F}$, $\bold{H}$, $\sigma$, and the state of each cluster---can be derived from symmetries in the network~\cite{pecora_ncomm2014,Golubitsky1985,Golubitsky2002}.

The symmetries of the network (whose structure is represented by $A$) are determined by the automorphism group ${\textrm {Aut}}(A)$, defined as the group formed by all node permutations that hold invariant the network topology [and thus Eq.~\eqref{pecora_rate}, which governs the dynamics]~\cite{Symmetry_complex,pecora_ncomm2014,godsil2013algebraic}.
Such a network-invariant permutation $i \rightarrow \pi(i)$ satisfies $A_{ij}=A_{\pi(i)\pi(j)}$.
The orbit $\varphi(G,i)$ of node $i$ under the group $G={\textrm {Aut}}(A)$ is defined as $\varphi(G,i):=\{\pi(i)\,|\,\pi \in G\}$, i.e., the set of all nodes to which node $i$ is mapped under all permutations in $G$.
Since $\varphi(G,i)=\varphi(G,j)$ holds for all nodes $j \in \varphi(G,i)$ by the group property of $G$, each node in the network belongs to a unique orbit of $G$.
Thus, the set of all orbits of $G$ defines a partition of the network into symmetry clusters, forming a candidate CS pattern.
However, and central to this study, there are potentially many other candidate CS patterns determined in the same way by any subgroup $G$ of ${\textrm {Aut}}(A)$~\cite{pecora_arxiv2015} [where below we use the term subgroup and the notation $G$ to refer either to ${\textrm {Aut}}(A)$ or any of its proper subgroups].
Figure~\ref{Fig:Independent_Intertwined_examples_schematic} shows an example of such a pattern.

For a given subgroup $G$ and the associated candidate CS pattern having clusters $C_1,\ldots,C_M$, there are generally multiple synchronous states respecting that pattern.
Each such state 
$\{\bold{s}_m(t)\}_{1\le m\le M}$, with $\bold{x}_i(t)=\bold{s}_m(t)$ for all $i\in C_m$ and for all $t$, satisfies 
\begin{equation}
\dot{\bold{s}}_m = {\bold F}(\bold{s}_m) + \sigma\sum_{m'=1}^M \widetilde{A}_{mm'}{\bold H}(\bold{s}_{m'}),
\label{eqn:reduced}
\end{equation}
where we have defined $\widetilde{A}_{mm'}:=\sum_{j \in C_{m'}} A_{ij}$ with $i\in C_m$.
This can be verified by substituting the synchronous state into Eq.~\eqref{pecora_rate} and rewriting the summation in Eq.~\eqref{pecora_rate} as $\sum_{m'=1}^M \sum_{j \in C_{m'}}$.
Note that $\widetilde{A}_{mm'}$ is properly defined because the invariance of $A$ under all permutations in $G$ (i.e., $A_{ij}=A_{\pi(i)\pi(j)}$ for all $\pi \in G$) can be used to show that $\widetilde{A}_{mm'}$ as defined does not depend on the choice of node $i$ within $C_m$.
The matrix $\widetilde{A} = (\widetilde{A}_{mm'})_{1\le m,m' \le M}$ can be interpreted as the (possibly directed weighted) adjacency matrix of a coarse-grained version of the original network (called a quotient network~\cite{stewart:609}), where $\widetilde{A}_{mm'}$ is the number of links from a node in $C_m$ to the cluster $C_{m'}$.
Note also that Eq.~\eqref{eqn:reduced} [thus the set of possible synchronous states for Eq.~\eqref{pecora_rate}]
is fully determined by the candidate CS pattern and does not directly depend on the associated subgroup $G$ (which may not be unique in general).
While similar candidate CS patterns and the corresponding synchronous states can also be formulated using symmetry groupoids~\cite{stewart:609} and external equitable partitions~\cite{EEP}, here we focus on those based on symmetry (sub)groups, as they facilitate our analysis below.

Whether a given candidate CS pattern can actually be observed in the system depends on whether the synchronization of the individual clusters is stable.
Different clusters are generally interrelated and, in particular, can have identical stability, which can be described by the notion of intertwined clusters~\cite{pecora_ncomm2014} in the special case where $G$ is the full automorphism group ${\textrm {Aut}}(A)$.
However, direct extension of this notion to an arbitrary subgroup $G$ does not lead to a consistent definition of intertwined clusters (Supplemental Material~\cite{SM}, Sec.~II).
Below we develop a comprehensive theory that overcomes this difficulty and fully describes the interrelationship between the synchronization stability of the clusters in a given candidate CS pattern.
This theory will provide a unique grouping of these clusters and the associated decoupling of the stability equations for clusters belonging to different groups.
Specifically, we will define a coordinate system in which the stability equations for each group of clusters (denoted $C_1,...,C_{M'}$, with sizes $c_1,\ldots,c_{M'}$, respectively) are coupled only within the group.  
The decoupled equations read:  
\begin{equation}
{\dot{\boldsymbol \eta}^{(m)}}_{\kappa}=D{\bold F}(\bold{s}_m){\boldsymbol \eta}^{(m)}_{\kappa}+
\sigma\sum_{m'=1}^{M'}\sum_{\kappa'=2}^{c_{m'}} B^{(mm')}_{\kappa\kappa'} D{\bold H}({\bold s}_{m'}){\boldsymbol \eta}^{(m')}_{\kappa'},
\label{Eq:pecora_variation_block}
\end{equation}
where ${\boldsymbol \eta}^{(m)}_2,\ldots,{\boldsymbol \eta}^{(m)}_{c_m}$ are variables that represent perturbations transverse to the synchronization manifold $\{ (\bold{x}_1,\ldots,\bold{x}_N) \,|\, \bold{x}_i=\bold{x}_j \text{ for all } i,j \in C_m \}$ of cluster $C_m$,
$D{\bold F}$ and $D{\bold H}$ are the Jacobian matrices of ${\bold F}$ and ${\bold H}$, respectively, $\{\bold{s}_m(t)\}$ is the synchronous state corresponding to the given CS pattern, and $B^{(mm')}_{\kappa\kappa'}$ is the coupling coefficient between ${\boldsymbol \eta}^{(m)}_{\kappa}$ and ${\boldsymbol \eta}^{(m')}_{\kappa'}$.
This means, in particular, that (a)~perturbations applied to a cluster in one group do not propagate to those in other groups, (b)~clusters in the same group must have the same stability, and (c)~clusters belonging to different groups can have different stability.
For a group with $M'=1$ (i.e., a single cluster), we will show that the coordinate system can be chosen so that Eq.~\eqref{Eq:pecora_variation_block} further reduces to
\begin{equation}
\dot{\boldsymbol\eta}_{\kappa}^{(m)}=\bigl[D{\bold F}(\bold{s}_m)+\sigma\lambda_{\kappa}^{(m)}D{\bold H}(\bold{s}_m)\bigr]{\boldsymbol\eta}_{\kappa}^{(m)},
\label{Eq:pecora_variation_eigenvector}
\end{equation}
where $\lambda^{(m)}_{\kappa}$ is an eigenvalue of $A$.
This generalizes the equation defining a master stability function for the stability analysis of complete synchronization (i.e., the special case $M=1$) in networks of diffusively coupled oscillators~\cite{PRL_Master}.

To define the grouping of the clusters $C_1,\ldots,C_M$ in a given candidate CS pattern, we first categorize the nontrivial clusters (i.e., those containing more than one node) into two types: those that are independently synchronizable and those that are not.
We say that $C_m$ is an {\it independently synchronizable cluster} (ISC) if the network has a state in which all nodes in $C_m$ are synchronized while none of the other nodes are required to be synchronized with any other nodes in the network.
Mathematically, such a cluster can be completely characterized by the following property of the network structure: there is a subgroup $G'$ of ${\textrm {Aut}}(A)$ for which ${\bold C}^{(G')}$ contains only the cluster $C_m$, where ${\bold C}^{(G')}$ denotes the set of all nontrivial clusters associated with $G'$.
The clusters $C_6,\ldots,C_9$ shown in Fig.~\ref{Fig:Independent_Intertwined_examples_schematic} are all ISCs.

What happens if $C_m$ is not an ISC?
In that case, we can still find a set of clusters (containing $C_m$) that is independently synchronizable as a whole, i.e., the network has a state in which all clusters in the set are synchronized (although the node states can be different for different clusters) while other parts of the network are not required to be synchronized.
Formally, we define an {\it ISC set} to be a minimal such set, i.e., a set ${\bold C}$ of nontrivial clusters satisfying the following conditions: 1)~there exists a subgroup $G'$ for which ${\bold C}={\bold C}^{(G')}$ (i.e., ${\bold C}$ is independently synchronizable), and 2) there is no subgroup $G''$ for which ${\bold C}^{(G'')}$ is a proper subset of ${\bold C}$ (i.e., ${\bold C}$ is a smallest such set).
For example, the clusters $C_1$ and $C_2$ in Fig.~\ref{Fig:Independent_Intertwined_examples_schematic} form an ISC set.
Our definition of ISC sets thus provides a higher-order organization of the network nodes into ``clusters of clusters''; for any given network structure and any candidate CS pattern [associated with some subgroup of ${\textrm {Aut}}(A)$], the (nontrivial) clusters can be uniquely grouped into ISCs and ISC sets~\cite{ref_com}.
We provide a proof of this unique grouping and also an efficient algorithm~\cite{software} for computing the grouping (Supplemental Material~\cite{SM}, Sec.~III).

We now construct a cluster-based coordinate system that leads to the decoupling in Eq.~\eqref{Eq:pecora_variation_block}.
For each cluster $C_{m}$, we first define a unit vector ${\bold u}^{(m)}_1$ parallel to the synchronization manifold for that cluster by setting its $i$th component to $1/\sqrt{c_m}$ if $i \in C_{m}$ and zero otherwise.
We then choose any set of vectors ${\bold u}^{(m)}_2,\ldots,{\bold u}^{(m)}_{c_m}$ that, together with ${\bold u}^{(m)}_1$, 
form an orthonormal basis for the $c_m$-dimensional subspace associated with the cluster $C_m$ (i.e., the subspace of the $N$-dimensional node coordinate space spanned by $\{ {\bold e}_i \}_{i \in C_m}$, where ${\bold e}_i$ denotes the vector in which the $i$th component is one and all others are zero).  It can be shown using the group-theoretical properties of the ISC sets that the corresponding similarity transformation is guaranteed to block-diagonalize the adjacency matrix $A$, with one diagonal block for each ISC or ISC set. 
Applying this transformation to the variational equations for an ISC set $C_1,...,C_{M'}$, we obtain Eq.~\eqref{Eq:pecora_variation_block}. 
In this equation,
the diagonal block of the transformed $A$ corresponding to the selected ISC set is further divided into smaller blocks $B^{(mm')}:=(B^{(mm')}_{\kappa\kappa'})$ representing the coupling between clusters $C_m$ and $C_{m'}$ within the same ISC set.
In contrast, between different ISC sets there is no coupling term in Eq.~\eqref{Eq:pecora_variation_block}; this indicates that the CS stability of one ISC set can be different from that of another ISC set.
We can further show that, within the same ISC set, there is no choice of a basis that would decouple the stability of different clusters. 
For ISCs (i.e., those for which $M'=1$), we can choose the basis vectors (except for ${\bold u}^{(m)}_1$) to be eigenvectors of $A$, which would diagonalize all the corresponding diagonal blocks of $A$ and decompose the variational equations into individual eigenmodes, which leads to Eq.~\eqref{Eq:pecora_variation_eigenvector}.
Similar decoupling of synchronization stability between different ISC sets can also be established for a general class of diffusively coupled oscillators (Supplemental Material~\cite{SM}, Sec.~IV for full details).

Our construction of the cluster-based coordinates (see Supplemental Material~\cite{SM}, Sec.~III C for an algorithm~\cite{software}) has significant computational advantage over 
the existing method of constructing block-diagonalizing coordinates based on irreducible representation (IRR) of subgroup $G$~\cite{pecora_ncomm2014} (detailed in Supplemental Material~\cite{SM}, Sec.~III A).
Figure~\ref{Fig:Intertwined_examples_schematic} shows the time it takes to compute the two coordinate systems for $G = {\textrm {Aut}}(A)$ as a function of $|{\textrm {Aut}}(A)|$ and $N$ using a single processor core on a workstation.
We observe that, as the network size and the number of symmetries grow, the computational time grows very quickly 
for the IRR coordinates (and the particular implementation fails to compute for $N>12$), while it grows much slower 
for the cluster-based coordinates 
and appears to saturate as a function of $|{\textrm {Aut}}(A)|$.

\begin{figure}[t!]
\includegraphics[width=1.0\linewidth]{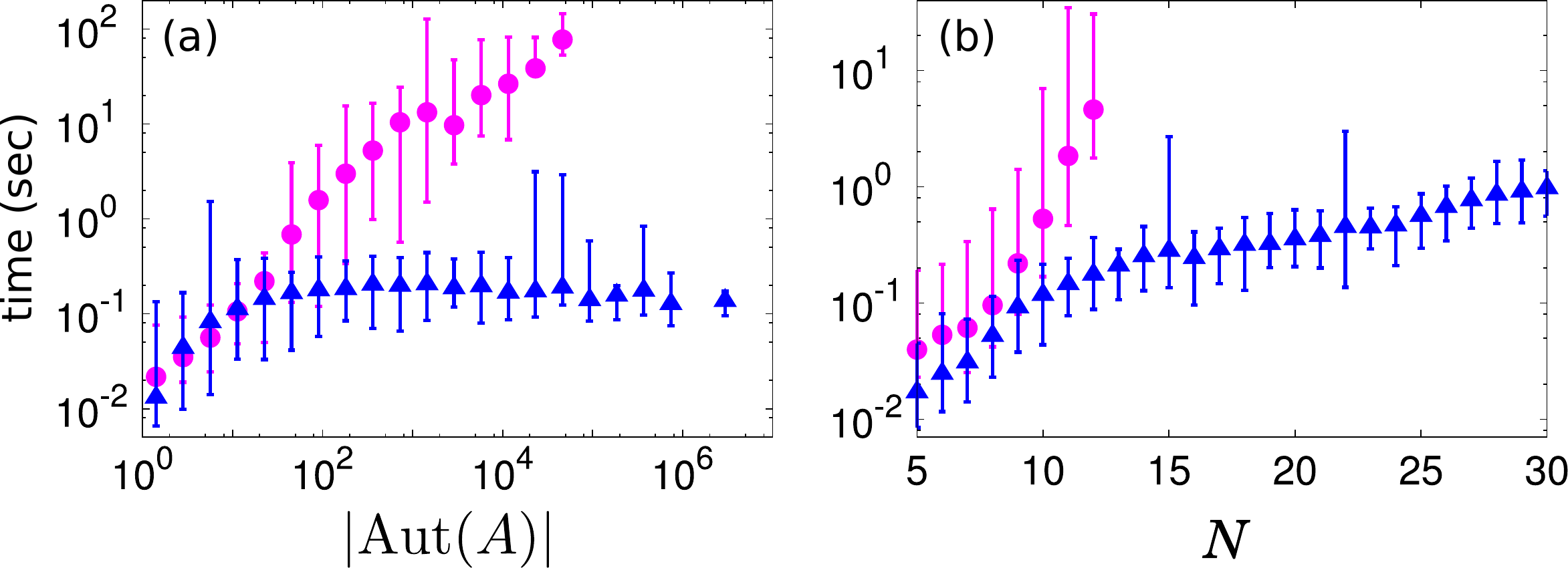} \vspace{-7mm}
\caption{
Computational time for constructing the cluster-based coordinates $(\blacktriangle)$ and the 
IRR coordinates $(\bullet)$ for Erd\H{o}s-R\'enyi random networks.
(a) Average CPU time vs.\ the number of symmetries $|{\textrm {Aut}}({A})|$, estimated from a sample of $10^3$ networks with $N=12$ nodes and $L$ links.
We vary $L$ in $N-1 \le L \le N(N-1)/2-2$, bin the resulting $|{\textrm {Aut}}({A})|$ values, and calculate the average CPU time in each bin.
The IRR algorithm fails for $|{\textrm {Aut}}({A})|>5 \times 10^4$.
(b) Average CPU time vs.\ the network size $N$, estimated from a sample of $100$ networks with $N$ nodes and $L=N(N-1)/2-5$ links.
The IRR algorithm fails for $N \geq 13$ because $|{\textrm {Aut}}({A})|$ becomes too large for such $N$.
For both (a) and (b), only connected networks are used, and error bars indicate the range of observed values.
}\label{Fig:Intertwined_examples_schematic}
\end{figure}

The theory we established above can be used to identify chimera states that are permanently stable, when applied to a fully symmetric network [i.e., one in which ${\textrm {Aut}}(A)$ has only one symmetry cluster].
To do this, we need to consider a proper subgroup of ${\textrm {Aut}}(A)$ that has at least one ISC strictly smaller than the network (since a chimera state requires at least one stable cluster and one unstable cluster).
This suggests the following two-step procedure for finding stable chimera states.
First, we choose a fully symmetric network structure that has an ISC $C_1$ that is strictly smaller than the network itself.
Then, we find system parameters satisfying the following conditions:
(i) the complete synchronization of the network [i.e., any state with $\bold{x}_i(t)=\bold{s}(t)$, $\forall i$ in Eq.~\eqref{pecora_rate}] is unstable;
(ii) for each candidate CS pattern in which $C_1$ appears, the synchronization of $C_1$ is stable but the synchronization of the other clusters is unstable according to Eqs.~\eqref{Eq:pecora_variation_block} and/or \eqref{Eq:pecora_variation_eigenvector}.
These conditions are necessary for a symmetry cluster-based chimera state to exist and persist indefinitely.

\begin{figure}[t!]
\includegraphics[width=1.0\linewidth]{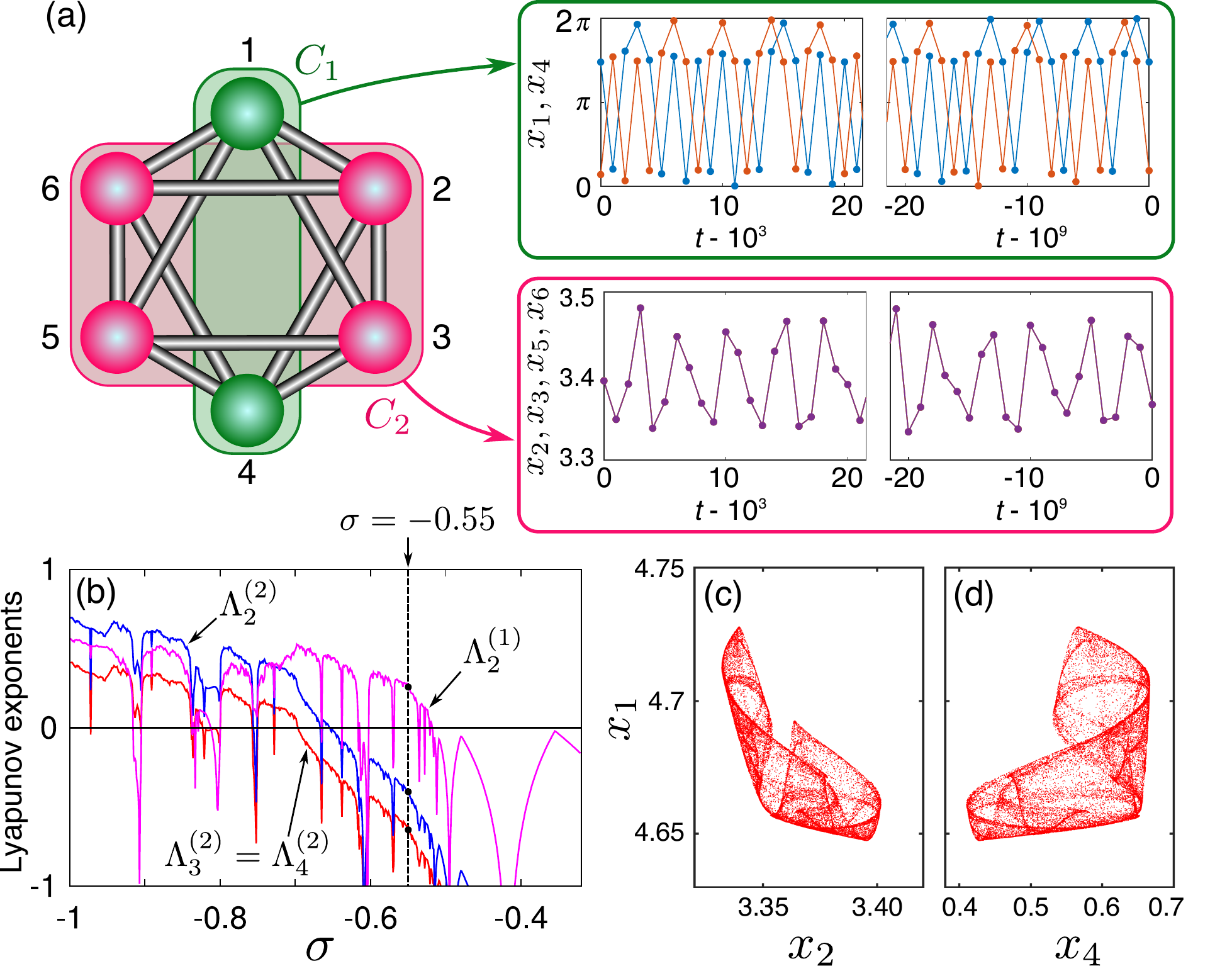}
\caption{
Permanently chimera state in the electro-optic system described by Eq.~\eqref{EO_rate}.
We use $\beta=\frac{2\pi}{3}-4\sigma$ and $\delta=\frac{\pi}{6}$ for a given coupling strength $\sigma$.
(a) Six-node ring structure of the network.
Shown in the two boxes on the right are time series of the nodes in clusters $C_1$ and $C_2$ obtained from directly iterating Eq.~\eqref{EO_rate} for $10^9$ time steps with $\sigma = -0.55$. 
(b) Transverse Lyapunov exponents for $C_1$ [$\Lambda^{(1)}_2$ (magenta)] and $C_2$ [$\Lambda^{(2)}_2$ (blue), $\Lambda^{(2)}_3=\Lambda^{(2)}_4$ (red)], estimated over $10^7 \leq t \leq 2 \times 10^7$.
(c--d) Last $10^5$ iterations in the time series, where we plot $x_1$ against $x_2$ (c) and $x_4$ (d), showing only every $4$th iterate (since the state is approximately periodic with period $4$).
We thus see that this is a chimera state in which $C_2$ is stably synchronized, while the dynamics within $C_1$ is chaotic.
}\label{Fig:Chimera_EO_regular_N_6_k_4}
\end{figure}

As an example of applying this approach, we consider the electro-optic system studied in Ref.~\cite{pecora_ncomm2014}, whose dynamics is governed by a discrete-time analog of Eq.~\eqref{pecora_rate}:
\begin{equation}
x_i(t+1)=\Big[\beta\mathcal{I}(x_i(t))+\sigma\sum_{j=1}^N A_{ij}\mathcal{I}(x_j(t))+\delta \Big]\ {\textrm {mod}}\ 2\pi,
\label{EO_rate}
\end{equation}
where $x_i(t)$ is the state variable for node $i$, the function $\mathcal{I}(x)=[1-{\mathrm {cos}}(x)]/2$ represents light intensity, and the parameter $\delta>0$ suppresses the trivial solution $x_i(t)\equiv 0$.
We use the six-node ring network shown in Fig.~\ref{Fig:Chimera_EO_regular_N_6_k_4}(a), which can be partitioned into two ISCs, $C_1 = \{1,4\}$ and $C_2 = \{2,3,5,6\}$.
For this system, condition~(i) can be verified to hold true if $\sigma < \frac{\pi}{9} - \frac{2}{3}$, $\beta = \frac{2\pi}{3}-4\sigma$, and $\delta = \frac{\pi}{6}$.
For condition~(ii), we only need to study the stability of states in which both $C_1$ and $C_2$ are synchronized, since no smaller nontrivial cluster within $C_1$ is possible.
Computing the transverse Lyapunov exponents for these clusters using a discrete-time analog of Eq.~\eqref{Eq:pecora_variation_eigenvector}, we find ranges of $\sigma$ for which the system satisfies condition~(ii) [Fig.~\ref{Fig:Chimera_EO_regular_N_6_k_4}(b)]:  $\Lambda^{(1)}_2>0$, $\Lambda^{(2)}_2<0$, and $\Lambda^{(2)}_3=\Lambda^{(2)}_4<0$, where $\Lambda^{(1)}_2$ denotes the exponent for $C_1$, while $\Lambda^{(2)}_2$, $\Lambda^{(2)}_3$, and $\Lambda^{(2)}_4$ denote the exponents for $C_2$.
Note that we always have $\Lambda^{(2)}_3=\Lambda^{(2)}_4$ for this system due to the degeneracy of the eigenvalues associated with these exponents.
Having verified both conditions (i) and (ii), the system is likely to exhibit a permanent chimera state.
Indeed, Fig.~\ref{Fig:Chimera_EO_regular_N_6_k_4} shows the trajectory of an example chimera state that emerges from a completely random initial condition and persists even after iterating Eq.~\eqref{EO_rate} for $10^9$ time steps.
In this state, the dynamics of node $1$ appears chaotic with respect to that of both node $2$ and node $4$ [Fig.~\ref{Fig:Chimera_EO_regular_N_6_k_4}(c--d)].
This provides further evidence that the system is in a permanent stable chimera state. 
We also identify permanent chimera states in larger networks using the same approach (see Supplemental Material~\cite{SM}, Sec.~V for details and for a larger network example). 

The theory of ``clusters of clusters'' we developed here answers a fundamental question about how the stability of synchronous clusters are interrelated.  Moreover, it provides a mechanism for a fully symmetric network to be in a permanently stable state that exhibits coherence and incoherence simultaneously.
Our formulation of network-structural conditions under which such chimera states are possible explain why some networks are more likely to exhibit chimeras than others; for example, the observed prevalence of chimera states in star networks~\cite{star_chimera} is due to the property that any partition of the end nodes yields a CS pattern in which all clusters are ISCs.
Our approach for finding chimera states in a fully symmetric network can also be applied to a given symmetry cluster in an arbitrary (not necessarily fully symmetric) network to identify sub-cluster chimeras: symmetry breaking that leads to the coexistence of coherent and incoherent dynamics within that cluster.
We hope our work will lead to the discovery of new patterns of synchronization that have not been anticipated before, and also stimulate further studies on cluster synchronization.

\begingroup
\renewcommand{\addcontentsline}[3]{}

\begin{acknowledgments}
This work was supported by ARO (Award No.~W911NF-15-1-0272) and by National Research Foundation of Korea (Grant No. NRF-2017R1C1B1004292).
\end{acknowledgments}


\endgroup 

\clearpage
\renewcommand{\baselinestretch}{1.1}
\setcounter{MaxMatrixCols}{20}
\onecolumngrid

\begin{center}
{\bf\Large Supplemental Material}\\ 
\bigskip
{\it Stable Chimeras and Independently Synchronizable Clusters}
\vskip 1.0mm

{Young Sul Cho, Takashi Nishikawa, and Adilson E. Motter}
\end{center}

\setcounter{equation}{0}
\setcounter{figure}{0}
\setcounter{table}{0}
\setcounter{page}{1}
\makeatletter
\renewcommand{\theequation}{S\arabic{equation}}
\renewcommand{\thefigure}{S\arabic{figure}}

\tableofcontents

\clearpage

\section{C\lowercase{lusters in the network of} F\lowercase{ig}. 1}

The candidate CS pattern shown in Fig.~1 is defined by the subgroup 
\begin{equation}
G=\bigl\langle(1,2), (3,4), (1,3)(2,4)(5,6), (8,9)(10,11)(12,13), (15,16), (18,19), (20,21), (22,23), (23,24)\bigr\rangle,
\end{equation} 
where $(1,2)$ denotes the permutation in which node $1$ is swapped with node $2$, while $(1,2)(3,4)$ denotes the permutation in which node $1$ is swapped with node $2$ as node $3$ is concurrently swapped with node $4$, and $\langle S\rangle$ denotes the subgroup generated by a subset $S$ of permutations.
For this pattern the unique grouping consists of four ISCs ($C_6,\ldots,C_9$) and two ISC sets ($\{C_1,C_2\}$ and $\{C_3,C_4,C_5\}$).
For example, $C_9$ is an ISC because ${\bold C}^{(G')} = \{C_9\}$ with $G' = \langle(22,23), (23,24)\rangle$, i.e., it is the only nontrivial cluster in the CS pattern associated with $G'$.
Similarly, $\{C_1,C_2\}$ is an ISC set because ${\bold C}^{(G')} = \{C_1,C_2\}$ with $G' = \langle (1,3)(2,4)(5,6), (1,2), (3,4) \rangle$.

\section{G\lowercase{eometric decomposition and intertwined clusters}}
\label{sec:geometric_fail}

\begin{figure}[h]
\includegraphics[width=0.3\linewidth]{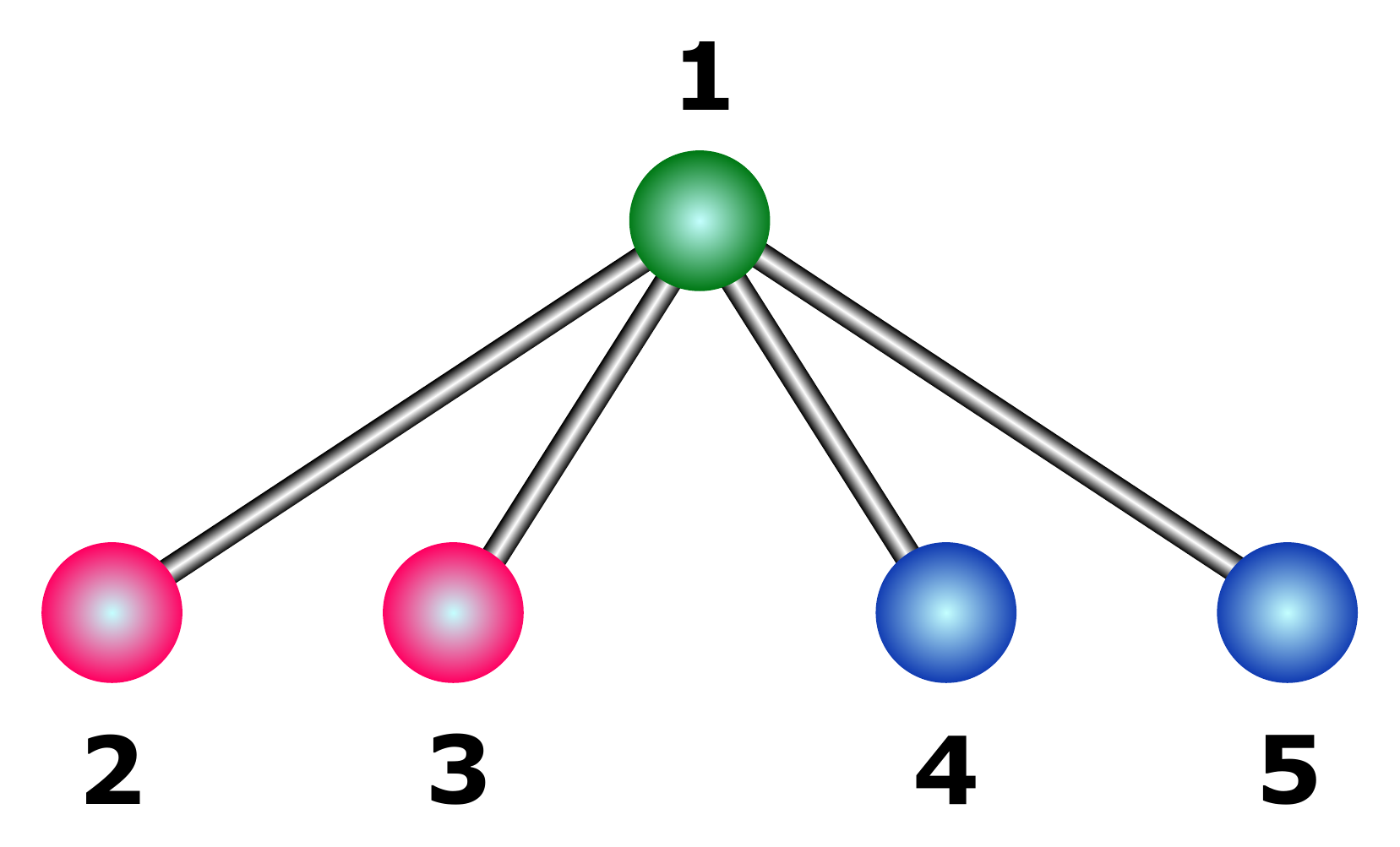}
\caption{Example network with two nontrivial clusters, $C_1=\{2,3\}$ and $C_2=\{4,5\}$, which can be classified either as intertwined or not intertwined depending on the choice a subgroup.}
\label{FigS:geometric_failure}
\end{figure}

The geometric decomposition of a group $G$ is defined to be a direct product decomposition, $G=H_1 \times \ldots \times H_{\nu} \times \ldots \times H_{\kappa}$, where $H_1,\ldots,H_{\kappa}$ are subgroups of $G$ whose support (the set of nodes moved by the permutations in the subgroup) do not overlap~\cite{Symmetry_complex, PRE_spectral}.
When $G={\textrm {Aut}}(A)$, two clusters $C_i$ and $C_j$ are said to be intertwined if they are both orbits of the same component $H_{\nu}$ from the geometric decomposition of ${\textrm {Aut}}(A)$~\cite{pecora_ncomm2014}.
Direct extension of this definition of intertwined clusters to arbitrary subgroups $G$ would lead to ambiguity, since whether clusters are intertwined or not would generally depend on the choice of $G$.
An example of such a situation is shown in Fig.~\ref{FigS:geometric_failure}.
On the one hand, the clusters $C_1=\{2,3\}$ and $C_2 =\{4,5\}$ would be classified as intertwined if the subgroup $G_1=\langle(2,3)(4,5)\rangle$ is used, since the geometric decomposition of $G_1$ has only one component, $H_1 = G_1$, and both clusters are orbits of $H_1$.
The permutation $(2,3)(4,5)$, the only permutation in $G_1$ other for the identity permutation, indeed moves the nodes in both clusters.
On the other hand, the two clusters would be classified as not intertwined if the subgroup $G_2=\langle(2,3),(4,5)\rangle$ is used, since the geometric decomposition of $G_2$ is $G_2 = H_1 \times H_2$, $H_1 = \langle(2,3)\rangle$, $H_2 = \langle(4,5)\rangle$, and $C_1$ is an orbit of $H_1$, while $C_2$ is not. 

\section{P\lowercase{roperties of} ISC\lowercase{s and} ISC \lowercase{sets}}

\subsection{IRR for the automorphism group and its subgroups}

For a given group $G$ (such as the automorphism group Aut($A$), i.e., the set of network-invariant node permutations), a {\it matrix representation} of $G$, which we denote here by $\Gamma$, is defined as a function that maps each element $\pi \in G$ to a matrix $\Gamma(\pi)$ and satisfies $\Gamma(\pi_1)\Gamma(\pi_2)=\Gamma(\pi_1\pi_2)$ for any $\pi_1,\pi_2 \in G$.
If there exists a matrix $T$ such that the similarity transformation $T\Gamma(\pi)T^{-1}$ has the same block diagonal form for all $\pi \in G$, this representation $\Gamma$ is called {\it reducible}.
This is because any one of the diagonal blocks defines a function mapping each $\pi\in G$ to a smaller matrix, serving as another matrix representation of $G$, thus decomposing the original representation $\Gamma$ into a combination of multiple representations.
If a representation is not reducible, it is called {\it irreducible}, and any given group $G$ has an associated set of IRRs, $\Gamma^{(1)},\ldots,\Gamma^{(R)}$.
By definition, any representation of $G$ can be decomposed into IRRs: there exists $T$ such that $T\Gamma(\pi)T^{-1}$ has the same block diagonal form for all $\pi \in G$ and each diagonal block is one of the $R$ irreducible representations.
A matrix $\Gamma^{(i)}(\pi)$ from the $i$-th IRR may appear as a block in multiple locations along the diagonal, and we use $a_i$ to denote the number of times it appears in this decomposition.
This decomposition can be expressed as a direct sum $\Gamma= \bigoplus_{i=1}^R a_i \Gamma^{(i)}$.
Here, $a_i$ is uniquely defined for a given representation $\Gamma$, irrespective of the choice of the similarity transformation $T$~\cite{tinkham}.

In the case $G$ is a group of automorphisms for a given $N$-node network [i.e., any subgroup of Aut($A$)], there is a special representation $\Gamma$ comprising the $N \times N$ {\it permutation matrices} $\Gamma(\pi)$, defined as $\Gamma(\pi)=[{\bold e}_{\pi(1)}, {\bold e}_{\pi(2)},\ldots, {\bold e}_{\pi(N)}]$, $\pi\in G$, where ${\bold e}_i$ is the unit column vector whose $i$th component equal to one and but all other components are zero.
We will use the notation $\Gamma$ for this special representation below.
Note that the relation $\Gamma(\pi)A=A\Gamma(\pi)$ is satisfied for each $\pi\in G$ because $\pi$ is an automorphism.
The transformation $T$ that decomposes this permutation matrix representation into IRRs can be found using the so-called {\it projection operators} defined by 
\begin{equation}\label{eqn:Pi_def}
P^{(i)}:=\frac{l_i}{|G|}\sum_{\pi\in G} \chi^{(i)}(\pi)^* \, \Gamma(\pi),
\end{equation}
where $l_i$ is the dimension of (the matrices in) the $i$-th IRR, $|G|$ is the cardinality of $G$, $\chi^{(i)}(\pi)$ is the trace of the $l_i$-dimensional matrices $\Gamma^{(i)}(\pi)$, and the star ($^*$) denotes the complex conjugate. 
These operators can be computed from the character table for the group $G$~\cite{tinkham}. 
For each $i$, the operator $P^{(i)}$ projects the $N$-dimensional node coordinate space onto the $n_i$-dimensional subspace associated with the $i$th IRR, where we define $n_i:=a_i l_i$.
This subspace can thus be fully characterized as the set of all vectors satisfying the relation
\begin{equation}\label{eqn:IRR_subspace}
P^{(i)}{\bold u}={\bold u}.
\end{equation}
We denote an arbitrary set of $n_i$ orthonormal vectors in this subspace by ${\bold t}^{(i)}_1,\ldots, {\bold t}^{(i)}_{n_i}$.
Such orthonormal basis can be computed from Eq.~\eqref {eqn:IRR_subspace} using, e.g., the singular value decomposition~\cite{pecora_ncomm2014}.
It is known that any two vectors belonging to the subspaces associated with different IRRs are orthogonal~\cite{tinkham}. 
Thus, the vectors ${\bold t}^{(1)}_1,\ldots,{\bold t}^{(1)}_{n_1},\ldots,{\bold t}^{(R)}_1,\ldots,{\bold t}^{(R)}_{n_R}$ form a basis for the entire $N$-dimensional space (since $\sum_{i=1}^R n_i=N$), and the similarity transformation that block diagonalizes matrices $\Gamma(\pi)$ can be constructed as 
\begin{equation}\label{eqn:T}
T=[{\bold t}^{(1)}_1,\ldots,{\bold t}^{(1)}_{n_1},\ldots,{\bold t}^{(R)}_1,\ldots,{\bold t}^{(R)}_{n_R}]^T.
\end{equation} 
This defines the {\it IRR coordinate system} associated with the given group $G$ for the given network.

\subsection{Block diagonalization of $A$ through cluster-based coordinate transformation}
\label{sec:EIG_Cluster_proof}

\begin{figure}[t]
\includegraphics[width=0.75\linewidth]{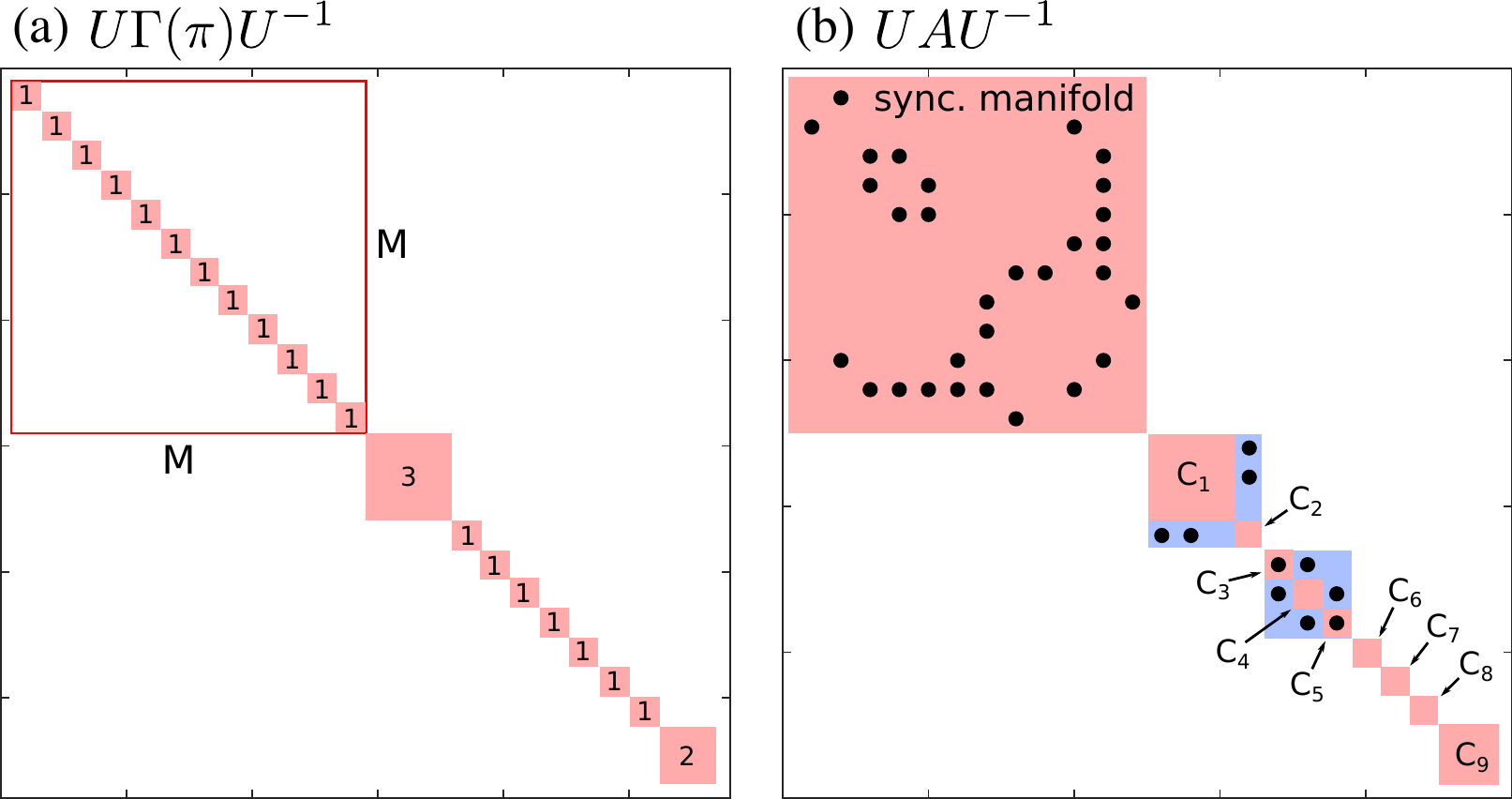}
\caption{Block diagonalization using the cluster-based coordinates, illustrated using the $24$-node example network and the CS pattern in Fig.~1 of the main text.
(a)~Block diagonal structure of $U\Gamma(\pi)U^{-1}$ common to all $\pi \in G$.
The first $M = 12$ (size-one) blocks correspond to the synchronization manifolds for the $M$ clusters, while the other $M'=9$ blocks correspond to the subspaces transverse to these manifolds for the $M'$ nontrivial clusters.
(b)~Block diagonal structure of $UAU^{-1}$.
The first block corresponds to the cluster synchronization manifold, while the other blocks correspond to ISCs and ISC sets (six blocks for the four ISCs and two ISC sets in this example). 
Black dots indicate nonzero values of the matrix components.}
\label{FigS:block_diag}
\end{figure}

For a given subgroup $G \subseteq {\textrm {Aut}}(A)$, the matrix that transforms into the cluster-based coordinate system defined in the main text is
\begin{equation}\label{eqn:U}
U=[{\bold{u}^{(1)}_{1}},\dots,{\bold{u}^{(M)}_{1}},{\bold{u}^{(1)}_{2}},\ldots,{\bold{u}^{(1)}_{c_1}},\ldots,{\bold {u}^{(M)}_{2}},\ldots,{\bold {u}^{(M)}_{c_M}}]^T.
\end{equation}
Since the basis vectors $\bold{u}^{(m)}_{\kappa}$ are chosen to be orthonormal, the matrix $U$ is orthogonal, i.e., satisfies $U^{-1}=U^T$.
The similarity transformation based on this matrix $U$ simultaneously block diagonalizes all permutation matrices $\Gamma(\pi)$, $\pi \in G$, i.e., $U\Gamma(\pi)U^{-1}$ is a block diagonal matrix of the form illustrated in Fig.~\ref{FigS:block_diag}(a) for all $\pi \in G$.
This is because, for each $m$, we have $\Gamma(\pi) \bold{u}^{(m)}_{1} = \bold{u}^{(m)}_{1}$ and $\Gamma(\pi) \bold{u}^{(m)}_{\kappa}$  is a linear combination of $\bold{u}^{(m)}_{2},\ldots,\bold{u}^{(m)}_{c_m}$ for any $\kappa \ge 2$ (both of which follow from the fact that any permutation $\pi\in G$ moves a node in a cluster $C_m$ only within that cluster).
This block diagonalization indicates that the representation $\Gamma$ is always reducible and that the $N$-dimensional space can be split into subspaces (one for each block) that are invariant under any permutations in $G$.
The first $M$ one-dimensional subspaces, spanned by ${\bold{u}^{(1)}_{1}},\dots,{\bold{u}^{(M)}_{1}}$, correspond to the trivial representation [and thus serves as the vectors ${\bold t}^{(1)}_1,\ldots,{\bold t}^{(1)}_{n_1}$ in Eq.~\eqref{eqn:T}].
In fact, these vectors span the entire subspace associated with the trivial representation, since a size-one diagonal block ($=1$) appears in $U\Gamma(\pi)U^{-1}$ if and only if the associated basis vector ${\bold u}$ satisfies $\Gamma(\pi) \bold{u} = \bold{u}$ for all $\pi \in G$, which we can show is possible only if ${\bold u}$ is a linear combination of the vectors ${\bold{u}^{(1)}_{1}},\dots,{\bold{u}^{(M)}_{1}}$.
We thus have $n_1 = M$.

We now show that the same transformation also block diagonalizes the adjacency matrix $A$ into a similar but slightly different form, illustrated in Fig.~\ref{FigS:block_diag}(b).
The first $M \times M$ block of $UAU^{-1}$ corresponds to the subspace spanned by ${\bold t}^{(1)}_m = {\bold u}^{(m)}_1$, $m=1,\ldots,M$, which represents the cluster synchronization manifold.
The other blocks correspond to the subspaces that are transverse to the synchronization manifold and are associated with ISCs and ISC sets.
To see that $UAU^{-1}$ has this block diagonal form (after re-indexing the clusters, if needed, so that those in the same ISC set appear consecutively), suppose first that $C_m$ is an ISC.
We will show that all components in the rows and columns corresponding to the basis vectors ${\bold u}^{(m)}_{2},\ldots,{\bold u}^{(m)}_{c_m}$ must be zero outside the $(c_m-1) \times (c_m-1)$ diagonal block.
From the definition of an ISC, there exists a subgroup $G'$ for which the only nontrivial cluster is $C_m$.
The IRR coordinates for this subgroup $G'$ has the form
\begin{equation}\label{eqn:T2}
\widetilde{T}=[\tilde{\bold t}^{(1)}_1,\ldots,\tilde{\bold t}^{(1)}_{\widetilde{M}},\tilde{\bold t}^{(2)}_1,\ldots,\tilde{\bold t}^{(2)}_{\tilde{n}_2},\ldots,\tilde{\bold t}^{(R)}_1,\ldots,\tilde{\bold t}^{(R)}_{\tilde{n}_R}]^T,
\end{equation} 
and the cluster-based coordinates has the form
\begin{equation}\label{eqn:U2}
\widetilde{U}=[\tilde{\bold u}^{(1)}_1,\ldots,\tilde{\bold u}^{(\widetilde{M})}_1,{\bold u}^{(m)}_2,\ldots,{\bold u}^{(m)}_{c_m}]^T,
\end{equation} 
where $\widetilde{M} := N-c_m+1$ is the number of clusters associated with subgroup $G'$, and we have chosen $\widetilde{U}$ so that $U$ and $\widetilde{U}$ share the same set of vectors, ${\bold u}^{(m)}_2,\ldots,{\bold u}^{(m)}_{c_m}$, associated with the cluster $C_m$.
Since $U$ and $\widetilde{U}$ both define orthonormal coordinate system for the same space, each vector in $U$ other than ${\bold u}^{(m)}_2,\ldots,{\bold u}^{(m)}_{c_m}$ can be written as a linear combination of the vectors $\tilde{\bold u}^{(1)}_1,\ldots,\tilde{\bold u}^{(\widetilde{M})}_1$.
The same argument we used above for $G$, $T$, and $U$ can now be applied to $G'$, $\widetilde{T}$, and $\widetilde{U}$ to see that $\widetilde{T}$ can be chosen so that $\tilde{\bold u}^{(\widetilde{m}')}_1 = \tilde{\bold t}^{(1)}_{\widetilde{m}'}$ for each $\widetilde{m}'$.
Thus, we can write 
\begin{equation}\label{eqn:umk1}
{\bold u}^{(m')}_{\kappa'} = \sum_{\widetilde{m}'=1}^{\widetilde{M}} c_1(\kappa', m', \widetilde{m}') \cdot \tilde{\bold u}^{(\widetilde{m}')}_1
= \sum_{\widetilde{m}'=1}^{\widetilde{M}} c_1(\kappa', m', \widetilde{m}') \cdot \tilde{\bold t}^{(1)}_{\widetilde{m}'}
\end{equation}
for all $m' \neq m$ with $\kappa' = 2,\ldots,c_{m'}$ and for all $m'$ with $\kappa'=1$,
where $c_1(\kappa', m', \widetilde{m}')$ are constant coefficients.
Also, since $\tilde{\bold u}^{(1)}_1,\ldots,\tilde{\bold u}^{(\widetilde{M})}_1$ are the only vectors that belong to the subspace associated with the trivial representation of $G'$, the vectors ${\bold u}^{(m)}_2,\ldots,{\bold u}^{(m)}_{c_m}$ in $\widetilde{U}$ must be associated with other IRRs.
Thus, these vectors can be expressed as linear combinations of the basis vectors corresponding to the nontrivial IRRs: 
\begin{equation}\label{eqn:umk2}
{\bold u}^{(m)}_{\kappa} = \sum_{i=2}^R \sum_{\widetilde{m}=1}^{\tilde{n}_i} c_2(\kappa, i, \widetilde{m}) \cdot \tilde{\bold t}^{(i)}_{\widetilde{m}}
\quad \text{for $\kappa=2,\ldots,c_m$.}
\end{equation}
Using Eqs.~\eqref{eqn:umk1} and \eqref{eqn:umk2}, the component of $UAU^{-1}$ in the row corresponding to the vector ${\bold u}^{(m)}_{\kappa}$ and the column corresponding to the vector ${\bold u}^{(m')}_{\kappa'}$ can be written as
\begin{equation}\label{eqn:off_diag_comp}
\begin{split}
({\bold u}^{(m)}_{\kappa})^T A {\bold u}^{(m')}_{\kappa'}
&= \Biggl[ \sum_{i=2}^R \sum_{\widetilde{m}=1}^{\tilde{n}_i} c_2(\kappa, i, \widetilde{m}) \cdot \tilde{\bold t}^{(i)}_{\widetilde{m}} \Biggr]^T \cdot A \cdot \Biggl[ \sum_{\widetilde{m}'=1}^{\widetilde{M}} c_1(\kappa', m', \widetilde{m}') \cdot \tilde{\bold t}^{(1)}_{\widetilde{m}'} \Biggr]\\
&= \sum_{i=2}^R \sum_{\widetilde{m}=1}^{\tilde{n}_i} \sum_{\widetilde{m}'=1}^{\widetilde{M}} 
c_2(\kappa, i, \widetilde{m}) \cdot c_1(\kappa', m', \widetilde{m}') \cdot 
\Bigl[ \bigl( \tilde{\bold t}^{(i)}_{\widetilde{m}} \bigr)^T \cdot A \cdot \tilde{\bold t}^{(1)}_{\widetilde{m}'} \Bigr],
\end{split}
\end{equation}
for all $m' \neq m$ with $\kappa=2,\ldots,c_m$ and $\kappa'=2,\ldots,c_{m'}$, as well as for all $m'$ with $\kappa=2,\ldots,c_m$ and $\kappa'=1$ (in which case this component is outside the diagonal block corresponding to the cluster $C_m$).
Let $\widetilde{P}^{(1)}$ denote the projection operator defined by Eq.~\eqref{eqn:Pi_def} for the trivial representation of $G'$.
Because each permutation matrix $\Gamma(\pi)$ commutes with $A$, so does $\widetilde{P}^{(1)}$ [which follows directly from Eq.~\eqref{eqn:Pi_def}].
Using this and applying Eq.~\eqref{eqn:IRR_subspace} to a basis vector $\bold{u} = {\bold t}^{(1)}_{\widetilde{m}'}$ for the subspace associated with the trivial representation, we have 
$\widetilde{P}^{(1)} A {\bold t}^{(1)}_{\widetilde{m}'} = A \widetilde{P}^{(1)}{\bold t}^{(1)}_{\widetilde{m}'} = A {\bold t}^{(1)}_{\widetilde{m}'}$, implying that the vector $A {\bold t}^{(1)}_{\widetilde{m}'}$ also belongs to the same subspace.
Since two vectors are known to be orthogonal if they belong to the subspaces associated with different IRRs, this leads to
$\bigl( \tilde{\bold t}^{(i)}_{\widetilde{m}} \bigr)^T \cdot A \tilde{\bold t}^{(1)}_{\widetilde{m}'} = 0$ whenever $i \neq 1$.
Thus, the component of $UAU^{-1}$ in Eq.~\eqref{eqn:off_diag_comp} is zero, since each term in the summation is zero.
A similar argument can be used to show that $({\bold u}^{(m')}_{\kappa'})^T A {\bold u}^{(m)}_{\kappa}$ is also zero.
For an ISC set, 
we use the subgroup $G'$ from condition~(1) in the definition of ISC sets to define $\widetilde{T}$ and $\widetilde{U}$.
A similar argument can then be made to show that 
$({\bold u}^{(m)}_{\kappa})^T A {\bold u}^{(m')}_{\kappa'}=({\bold u}^{(m')}_{\kappa'})^T A {\bold u}^{(m)}_{\kappa} = 0$, where $m$ indexes the clusters in the ISC set and $m'$ the other clusters.
Altogether, we have shown that all components of $UAU^{-1}$ outside the diagonal blocks are zero, as illustrated in Fig.~\ref{FigS:block_diag}.

Next we show that cluster-based coordinates cannot further block diagonalize those blocks associated with ISC sets.
Let ${\bold C} := \{C_1,\ldots,C_{M'}\}$ denote an arbitrary ISC set (after suitable re-indexing of the clusters).
It is sufficient to show that the corresponding block cannot be transformed into a block diagonal form with two blocks, one corresponding to the set of clusters ${\bold C}_1 := \{ C_1,\ldots,C_{M''} \}$ and the other to the set ${\bold C}_2 := \{ C_{M''+1},\ldots,C_{M'} \}$, by any choice of the basis vectors ${\bold{u}}^{(m)}_{\kappa}$, $1 \le m \le M'$, $2 \le \kappa \le c_m$.
We prove this by contradiction.
Suppose there is a choice of matrix $U$ achieving such further block diagonalization.
We will show that this implies the existence of a subgroup $G'' \subseteq {\textrm {Aut}}(A)$ satisfying ${\bold C}^{(G'')} = {\bold C}_1$, which violates
the assumption that ${\bold C}$ is an ISC set (specifically, the second condition in the definition of an ISC set).
We first define $G''$ to be the set of all permutations that can be obtained from restricting node movements of $\pi \in G$ to within the clusters in ${\bold C}_1$ (which is possible because no $\pi \in G$ moves nodes between $\bold{C}_1$ and $\bold{C}_2$).
Since $G''$ is just a restriction of $G$ to a subset of nodes, it is straightforward to show that it satisfies all the requirement to be a group (the closure and associativity properties, as well as the existence of the identity and the inverse elements) and that ${\bold C}^{(G'')} = {\bold C}_1$ (i.e., that the set of all nontrivial clusters in the CS pattern generated by $G''$ is exactly ${\bold C}_1$).
We are thus left to show that $G'' \subseteq {\textrm {Aut}}(A)$.
Let $\pi \in G$ be the permutation from which $\pi'' \in G''$ is obtained by restricting to ${\bold C}_1$.
Since $\Gamma(\pi'')$ is the permutation matrix obtained from $\Gamma(\pi)$ by moving all $1$'s in the rows not corresponding to ${\bold C}_1$ to the diagonal, $U\Gamma(\pi'') U^{-1}$ is obtained from $U\Gamma(\pi) U^{-1}$ by replacing the diagonal blocks not corresponding to ${\bold C}_1$ by identity matrices of suitable sizes.
From the relation $\Gamma(\pi) A = A \Gamma(\pi)$, we have 
\begin{equation}
U\Gamma(\pi) U^{-1} \cdot U A U^{-1} = U A U^{-1} \cdot U \Gamma(\pi) U^{-1},
\end{equation}
i.e., the matrices $U\Gamma(\pi) U^{-1}$ and $U A U^{-1}$ commute.
This commutativity property actually holds for individual diagonal blocks of $U A U^{-1}$ because each diagonal block of $U\Gamma(\pi) U^{-1}$ is contained in a diagonal block of $U A U^{-1}$ (see Fig.~\ref{FigS:block_diag}).
Due to the structure of $U\Gamma(\pi'') U^{-1}$ just mentioned, we have
\begin{equation}
U\Gamma(\pi'') U^{-1} \cdot U A U^{-1} = U A U^{-1} \cdot U \Gamma(\pi'') U^{-1},
\end{equation}
and hence
\begin{equation}
\Gamma(\pi'') A = A \Gamma(\pi''),
\end{equation}
implying that $\pi''$ is an automorphism.
Since this argument is valid for an arbitrary $\pi'' \in G''$, we have $G'' \subseteq {\textrm {Aut}}(A)$ and thus conclude that $G''$ is a subgroup of ${\textrm {Aut}}(A)$ satisfying ${\bold C}^{(G'')} = {\bold C}_1$, contradicting the initial assumption that ${\bold C}$ is an ISC set.
Therefore, the diagonal block in $U A U^{-1}$ corresponding to ${\bold C}$ cannot be further block diagonalized with smaller diagonal blocks representing proper subsets of ${\bold C}$.

Finally, we note that the basis vectors $\bold{u}^{(m)}_{\kappa}$ with $\kappa \ge 2$ corresponding to ISCs can be chosen to be eigenvectors of $A$, since the network is assumed to be undirected.
This is because $UAU^{-1} = UAU^T$ is symmetric (since $A$ is symmetric), and this makes each block of $UAU^{-1}$ also symmetric and thus diagonalizable.
For a given ISC $C_m$, we can use the eigenvectors of the corresponding diagonal block of $UAU^{-1}$ to construct eigenvectors of $A$ that are linear combinations of ${\bold u}^{(m)}_2,\ldots,{\bold u}^{(m)}_{c_m}$ and span the subspace transverse to the synchronization manifold of $C_m$. 
These eigenvectors can thus be used as the basis vectors to replace ${\bold u}^{(m)}_2,\ldots,{\bold u}^{(m)}_{c_m}$, which would fully diagonalize the diagonal block and make the corresponding eigenvectors appear on the diagonal of that block.

\subsection{Algorithm for computing the coordinate transformation}
\label{sec:algorithm_U}

In this section we describe how to compute the cluster-based coordinate transformation matrix $U$ in Eq.~\eqref{eqn:U} for a given (undirected) network (with adjacency matrix $A$) and a given candidate CS pattern (with clusters $C_1,\ldots,C_M$).
The first $M$ vectors are computed directly from their definition in the main text (i.e., the $i$th component of ${\bold u}^{(m)}_1$ equals $1/\sqrt{c_m}$ if $i \in C_{m}$ and zero otherwise).
For each cluster $C_m$, the corresponding (transverse) vectors ${\bold u}^{(m)}_2,\ldots,{\bold u}^{(m)}_{c_m}$ are constructed from an initial set of (linearly independent) vectors, composed of ${\bold u}^{(m)}_1$ and the unit vectors ${\bold e}_i$, $i\in C_m$ (where the $i$th component of ${\bold e}_i$ is one while all the other components are zero).
We then replace the vectors ${\bold e}_i$ with new vectors ${\bold u}^{(m)}_{\kappa}$ using the Gram-Schmidt process, which ensures that ${\bold u}^{(m)}_1,\ldots,{\bold u}^{(m)}_{c_m}$ form an orthonormal set of vectors that span the subspace associated with cluster $C_m$.
This yields a matrix $U$ that meets all the requirements, except possibly the need of re-indexing the clusters.
We check if we need to re-index, we search for a nonzero off-diagonal block associated with two different clusters.
If there is such a block, we re-index the clusters to make those two clusters appear consecutively and re-define $U$.
Repeating this until there is no such block, we have obtain $U$ that block diagonalizes $A$ as described in Sec.~\ref{sec:EIG_Cluster_proof}.
If $C_m$ is an ISC, we can replace ${\bold u}^{(m)}_2,\ldots,{\bold u}^{(m)}_{c_m}$ with eigenvectors to diagonalize the corresponding block $B_m$ in $B = UAU^{-1}$.
Computing the eigenvectors of $B_m$, we obtain an orthogonal matrix  $V = (V_{\kappa\kappa'})_{2 \le \kappa, \kappa' \le c_m}$ for which
\begin{equation}
V B_m V^T = \begin{bmatrix}
\, \lambda^{(m)}_2 \!\!\! & & \\[-5pt]
 & \ddots & \\[-5pt]
 & & \lambda^{(m)}_{c_m}
\end{bmatrix}.
\end{equation}
We then replace ${\bold u}^{(m)}_{\kappa}$ by ${\bold v}^{(m)}_{\kappa} := \sum_{\kappa'=2}^{c_m} V_{\kappa\kappa'} {\bold u}^{(m)}_{\kappa'}$, which can be shown to be an eigenvector of $A$.
Thus, with this replacement for each ISC, the matrix $U$ diagonalizes all the diagonal blocks of $A$ corresponding to ISCs.

We provide~\cite{software} an implementation of the algorithm above in Python (filename: \verb|grouping_clusters.py|) as well as a convenient software tool (filename: \verb|all_CS_patterns.sage|) for choosing a candidate CS pattern valid for a given network, implemented using SageMath~\cite{sagemath}.
For a given $A$, we first obtain a set of generators for the automorphism group, $\textrm{Aut}(A)$.
We then partition this set into support-disjoint subsets, where we recall that the support of a set of generators is defined to be the set of nodes that move under some generator in the set.
This partition defines the geometric decomposition, $\textrm{Aut}(A) = H_1 \times \cdots \times H_{\nu} \times \cdots \times H_K$, 
where the subgroup $H_{\nu}$ is generated by the $\nu$th subset of generators in the partition. 
It is called ``geometric'' since it also partitions the network into non-overlapping subsets of nodes (i.e., the support of $H_{\nu}$'s), and the nodes in each set move only under the permutations in one specific subgroup $H_{\nu}$.
We then use SageMath to compute a list of all subgroups of each $H_{\nu}$, which together determine all possible partitions of the support of $H_{\nu}$ into clusters.
By choosing one partition for each $H_{\nu}$ and combining them, we obtain a valid candidate CS pattern.

\subsection{Unique grouping of clusters into ISCs and ISC sets}
\label{sec:Uniqueness_partition}

We first show that, for any given subgroup $G$ and the associated candidate CS pattern, we can always partition the set of all nontrivial clusters, ${\bold C}_0 := {\bold C}^{(G)} = \{ C_1,\ldots,C_{M'} \}$, into ISCs and ISC sets.
To do this, we use the following recursive procedure to define a sequence of partitions of ${\bold C}_0$, denoted ${\mathcal P}_1, {\mathcal P}_2,\ldots,$
where each partition ${\mathcal P}_i$ consists of disjoint subsets of ${\bold C}_0$ satisfying condition~(1) in the definition of ISC sets:
\begin{enumerate}
\item Define ${\mathcal P}_1$ to be the partition consisting of only the set ${\bold C}_0$ [which satisfies condition~(1) because ${\bold C}^{(G)}={\bold C}_0$].
\item For each $i=2,3,\ldots,$ define ${\mathcal P}_{i}$ from ${\mathcal P}_{i-1}$ by applying the following to each set in ${\mathcal P}_{i-1}$ [which is assumed to satisfy condition~(1)]:
\begin{enumerate}
 \item If the set additionally satisfies condition~(2) or contains only one cluster, then include it in ${\mathcal P}_{i}$.
 \item If the set does not satisfy condition~(2), then split it into two disjoint subsets that satisfy condition~(1) and include both in ${\mathcal P}_{i}$.  (We will show below that this is always possible.)
\end{enumerate}
\item Repeat step 2 until all sets in ${\mathcal P}_{i}$ satisfy one of the two conditions in step 2(a).
\end{enumerate}
Since the number of clusters in ${\bold C}_0$ is finite, this procedure will eventually stop.
The partition obtained at the end will consist only of ISCs and ISC sets, since in addition to satisfying condition~(1) each set in this partition will either additionally satisfies condition~(2) (in which case it is an ISC set) or contain only one cluster [which must be an ISC because the set satisfies condition~(1)].

To see why we can always perform step 2(b) above, suppose that ${\bold C}$ satisfying condition~(1) but not condition~(2).
This implies that there is a subgroup $G_1$ for which ${\bold C}_1 := {\bold C}^{(G_1)}$ is a proper subset of ${\bold C}$.
The subset ${\bold C}_1$ thus satisfies condition~(1).
We are left with showing that the subset ${\bold C}_2 := {\bold C} - {\bold C}_1$ of remaining clusters also satisfy condition~(1).
We apply to the set ${\bold C}_1$ the same argument we applied to an ISC set in Sec.~\ref{sec:EIG_Cluster_proof} in showing block diagonalization of $A$, noting that the argument relies only the fact that the ISC set satisfies condition~(1) and not condition~(2).
It follows that the components in the rows and columns of $UAU^{-1}$ corresponding to ${\bold C}_1$ but outside the diagonal block are all zero.
In particular, the diagonal block corresponding to ${\bold C}$ is block diagonalized with smaller diagonal blocks corresponding to ${\bold C}_1$ and ${\bold C}_2$.
This implies that we can now apply the argument from the same section for the impossibility of further block diagonalization of the block associated with an ISC set.
It follows that the group $G_2$ obtained by restricting $G$ to node movements within ${\bold C}_2$ satisfies $G_2 \subseteq {\textrm {Aut}}(A)$ and ${\bold C}^{(G_2)} = {\bold C}_2$, and thus satisfies condition~(1).
This shows that step 2(b) above can always be performed.

We now show the uniqueness of the partition obtained by the procedure above by contradiction.
Suppose we have two different partitions.
This implies that there are two different but overlapping ISC sets within ${\bold C}_0$, which can be expressed as ${\bold C}_1 \cup {\bold C}_2$ and ${\bold C}_2 \cup {\bold C}_3$, where ${\bold C}_i$, $i=1,2,3$ are disjoint, non-empty subsets of ${\bold C}_0$.
Consider the diagonal block of $UAU^{-1}$ using the cluster-based coordinates in Eq.~\eqref{eqn:U} after re-indexing the clusters so that ${\bold C}_1, {\bold C}_2, {\bold C}_3$ appear consecutively (in that order).
Since ${\bold C}_1 \cup {\bold C}_2$ is an ISC set, the argument in Sec.~\ref{sec:EIG_Cluster_proof} can be used to see that the diagonal sub-block corresponding to ${\bold C}_1 \cup {\bold C}_2$ is separate from that corresponding to ${\bold C}_3$.
Similarly, the diagonal sub-block corresponding to ${\bold C}_2 \cup {\bold C}_3$ is separate from that corresponding to ${\bold C}_1$.
Putting these two assertions together, we see that the diagonal sub-blocks corresponding to all three sets are actually separate from each other.
This implies that we can use the argument in Sec.~\ref{sec:EIG_Cluster_proof} to show that there exists a subgroup $G_2 \subseteq {\textrm {Aut}}(A)$ such that ${\bold C}^{(G_2)} = {\bold C}_2$.
This violates the condition~(2) for ${\bold C}_1 \cup {\bold C}_2$ to be an ISC set, which is a contradiction.

\subsection{Efficient algorithm to identify the unique grouping}

The recursive procedure described in Sec.~\ref{sec:Uniqueness_partition} provides an algorithm for finding the unique grouping of the clusters into ISCs and ISC sets, but it is inefficient because step $2$ requires searching through all possible subgroups of $\textrm{Aut}(A)$, which is computationally very expensive.
A better approach is to use the efficient construction of the matrix $U$ and its ability to block diagonalize $A$.
To find the unique grouping, we simply need to construct $U$ following the algorithm described in Sec.~\ref{sec:algorithm_U}, which identifies an ISC set as a set of clusters associated with the same diagonal block in $B = U A U^{-1}$.
Our Python implementation of the cluster-based coordinate algorithm~\cite{software} (filename: \verb|grouping_clusters.py|) also computes the unique grouping using this approach.

\section{C\lowercase{luster-based coordinate system and stability analysis}}

\subsection{Decoupling of stability equations}
\label{sec:Eq(3-4)}

For a given candidate CS pattern $C_1,\ldots,C_M$ (with $C_1,\ldots,C_{M'}$ being the nontrivial clusters), the variational equation of Eq.~(1) in the main text around a synchronous state ${\bold s}_m$ satisfying Eq.~(2) is given by
\begin{equation}
\delta\dot{\bold{x}}_i = D{\bold F}(\bold{s}_{m})\delta\bold{x}_i+\sigma\sum_{m'=1}^M\sum_{j \in C_{m'}} A_{ij}D{\bold H}(\bold{s}_{m'})\delta\bold{x}_j,
\label{pecora_rate_variation}
\end{equation}
for each $i \in C_m$, $m=1,\ldots,M$, where $\delta {\bold x}_i := {\bold x}_i - {\bold s}_m$ is the deviation from the synchronous state for node $i$.
Changing into the cluster-based coordinates defined in the main text, we obtain new variables,
${\boldsymbol \eta}^{(m)}_{\kappa} := \sum_{i \in C_{m}} u^{(m)}_{\kappa i}\delta {\bold x}_i$, 
$m=1,\ldots,M$, $\kappa = 1,\ldots,c_{m}$, 
where $u^{(m)}_{\kappa i}$ is the $i$-th component of the basis vector ${\bold u}^{(m)}_{\kappa}$ (which is nonzero only for $i \in C_m$).
Using the fact that $U$ in Eq.~\eqref{eqn:U} is an orthogonal matrix, we can invert this relation and write 
$\delta {\bold x}_i = \sum_{\kappa=1}^{c_m} u^{(m)}_{\kappa i}{\boldsymbol \eta}^{(m)}_{\kappa}$, where $m$ is the index of the cluster to which node $i$ belong (i.e., $i \in C_{m}$).
From the same fact, we also have the orthonormality relation $\sum_{i \in C_m} u^{(m)}_{\kappa i}u^{(m')}_{\kappa' i} = ({\bold u}^{(m)}_{\kappa})^T {\bold u}^{(m')}_{\kappa'} = \delta_{mm'}\delta_{\kappa\kappa'}$.
Using these relations, we rewrite Eq.~\eqref{pecora_rate_variation} in terms of the variables ${\boldsymbol \eta}^{(m)}_{\kappa}$:
\begin{equation}\label{derivation_pecora_variation_block}
\begin{split}
\dot{\boldsymbol{\eta}}^{(m)}_{\kappa}
&= \sum_{i \in C_m}u^{(m)}_{\kappa i} \delta \dot{{\bold x}}_i 
= \sum_{i \in C_m} u^{(m)}_{\kappa i}\Bigl[D{\bold F}({\bold s}_m)\delta {\bold x}_i + \sigma \sum_{m'=1}^M\sum_{j \in C_{m'}}A_{ij}D{\bold H}({\bold s}_{m'})\delta {\bold x}_j\Bigr] \\
&= D{\bold F}({\bold s}_m)\sum_{\kappa'=1}^{c_m}\sum_{i \in C_m} u^{(m)}_{\kappa i}u^{(m)}_{\kappa' i}{\boldsymbol \eta}^{(m)}_{\kappa'}
+ \sigma\sum_{m'=1}^M D{\bold H}({\bold {s}}_{m'}) \sum_{\kappa'=1}^{c_{m'}} \sum_{i \in C_m}\sum_{j \in C_{m'}} u^{(m)}_{\kappa i}A_{ij}u^{(m')}_{\kappa' j}{\boldsymbol \eta}^{(m')}_{\kappa'} \\
&= D{\bold F}({\bold s}_m){\boldsymbol \eta}^{(m)}_{\kappa}+\sigma\sum_{m'=1}^MD{\bold H}({\bold s}_{m'})\sum_{\kappa'=1}^{c_{m'}}B^{(mm')}_{\kappa\kappa'}{\boldsymbol \eta}^{(m')}_{\kappa'} \\
&= D{\bold F}(\bold{s}_m){\boldsymbol \eta}^{(m)}_{\kappa}+\sigma\sum_{m'=1}^{M'}\sum_{\kappa'=2}^{c_{m'}} B^{(mm')}_{\kappa\kappa'} D{\bold H}({\bold s}_{m'}){\boldsymbol \eta}^{(m')}_{\kappa'}
\end{split}
\end{equation}
for $m=1,\ldots,M$ and $\kappa = 2,\ldots,c_m$ (corresponding to the perturbations transverse to the cluster synchronization manifold), where we used the definition 
\begin{equation}
B^{(mm')}_{\kappa\kappa'}
:= ({\bold u}^{(m)}_{\kappa})^T A {\bold u}^{(m')}_{\kappa'}
= \sum_{i \in C_m}\sum_{j \in C_{m'}} u^{(m)}_{\kappa i}A_{ij}u^{(m')}_{\kappa' j},
\end{equation}
the result from Sec.~\ref{sec:EIG_Cluster_proof} that $B^{(mm')}_{\kappa 1} = 0$ for all $m'=1,\ldots,M$ [see Eq.~\eqref{eqn:off_diag_comp} and the argument following it], and the fact that for $m'>M'$ (i.e., for a trivial cluster) we have $c_{m'}=1$ and thus the summation $\sum_{\kappa'=2}^{c_{m'}}$ in Eq.~\eqref{derivation_pecora_variation_block} does not exist.
This proves Eq.~(3) in the main text.
The results from Sec.~\ref{sec:EIG_Cluster_proof} also implies that $B^{(mm')}_{\kappa\kappa'} = 0$ whenever $m$ and $m'$ correspond to different ISCs or ISC sets, and that no choice of coordinates can make $B^{(mm')}_{\kappa\kappa'}$ all zero for $m$ and $m'$ corresponding to different clusters within the same ISC set.
These, in the context of Eq.~\eqref{derivation_pecora_variation_block}, establish our claim that the synchronization stability of clusters belonging to different ISCs and ISC sets are decoupled, while those within the same ISC set cannot be. 

When the basis vectors $\bold{u}^{(m)}_{\kappa}$ with $\kappa \ge 2$ corresponding to ISCs are chosen to be eigenvectors of $A$, as was shown to be possible in Sec.~\ref{sec:EIG_Cluster_proof}, we have
$B^{(mm)}_{\kappa\kappa'} = \lambda^{(m)}_{\kappa} \delta_{\kappa\kappa'}$, $\kappa=2,\ldots,c_m$, where $\lambda^{(m)}_{\kappa}$ denotes the eigenvalue associated with the eigenvector $\bold{u}^{(m)}_{\kappa}$. 
Noting that in this case $B^{(mm')}_{\kappa\kappa'} = 0$ holds true whenever $m' \neq m$, Eq.~\eqref{derivation_pecora_variation_block} becomes
\begin{equation}
\begin{split}
\dot{\boldsymbol{\eta}}^{(m)}_{\kappa}
&= D{\bold F}(\bold{s}_m){\boldsymbol \eta}^{(m)}_{\kappa}
+\sigma\sum_{\kappa'=2}^{c_{m}} B^{(mm)}_{\kappa\kappa'} D{\bold H}({\bold s}_{m}){\boldsymbol \eta}^{(m)}_{\kappa'} \\
&= \Big[D{\bold F}({\bold s}_m)+\sigma\lambda^{(m)}_{\kappa}D{\bold H}({\bold s}_m)\Big]{\boldsymbol \eta}^{(m)}_{\kappa},
\end{split}
\end{equation}
thus establishing Eq.~(4) in the main text.

\subsection{Extension to networks of diffusively coupled oscillators}
\label{sec:Eq(7-8)}

Our results on the decoupling of cluster synchronization stability can be extended to the general class of systems with diffusive coupling (including Kuramoto-like systems) whose dynamics is governed by
\begin{equation}
\dot{{\bold x}}_i(t) = {\bold F}({\bold x}_i(t)) + \sigma\sum_{j=1}^NA_{ij}{\bold H}({\bold x}_j(t)-{\bold x}_i(t)).
\label{diffusive_rate}
\end{equation}
For a synchronous state ${\bold s}_m$ associated with a given candidate CS pattern $C_1,\ldots,C_M$ (with $C_1,\ldots,C_{M'}$ being the nontrivial clusters), the equation analogous to Eq.~(2) in the main text is
\begin{equation}
\dot{{\bold s}}_m = {\bold F}({\bold s}_m) + \sigma\sum_{m'=1}^{M}\widetilde{A}_{mm'}{\bold H}({\bold s}_{m'}-{\bold s}_m).
\label{eqn:diffusive_reduced}
\end{equation}
The variational equations of Eq.~\eqref{diffusive_rate} around the synchronous state is 
\begin{equation}
\begin{split}
\delta \dot{{\bold x}}_i &= D{\bold F}({\bold s}_{m}){\delta {\bold x}_i}-\sigma \sum_{m'=1}^M\sum_{k \in C_{m'}} A_{ik}{D{\bold H}}({\bold s}_{m'}-{\bold s}_{m})\delta {\bold x}_i + \sigma\sum_{m'=1}^M\sum_{j\in C_{m'}}A_{ij}{D{\bold H}}({\bold s}_{m'}-{\bold s}_{m})\delta {\bold x}_j \\
&= D{\bold F}({\bold s}_{m}){\delta {\bold x}_i}-\sigma\sum_{m'=1}^M\widetilde{A}_{mm'}{D{\bold {H}}}({\bold s}_{m'}-{\bold s}_{m})\delta{\bold x}_i+\sigma\sum_{m'=1}^M\sum_{j \in C_{m'}}A_{ij}{D{\bold H}}({\bold s}_{m'}-{\bold s}_{m})\delta{\bold x}_j.
\end{split}
\label{diffusive_variation}
\end{equation}
Using the cluster-based coordinate transformation and following the derivation in Sec.~\ref{sec:Eq(3-4)}, we rewrite Eq.~\eqref{diffusive_variation} as
\begin{equation}
\begin{split}
\dot{\boldsymbol \eta}^{(m)}_{\kappa}
&= \sum_{i \in C_m}u^{(m)}_{\kappa i} \delta \dot{{\bold x}}_i \\ 
&= \sum_{i \in C_m} u^{(m)}_{\kappa i}\biggl[D{\bold F}({\bold s}_{m}){\delta {\bold x}_i}-\sigma\sum_{m'=1}^M
\widetilde{A}_{mm'}{D{\bold {H}}}({\bold s}_{m'}-{\bold s}_{m})\delta{\bold x}_i+\sigma\sum_{m'=1}^M\sum_{j \in C_{m'}}A_{ij}{D{\bold H}}({\bold s}_{m'}-{\bold s}_{m})\delta{\bold x}_j\biggl] \\
&= D{\bold F}({\bold s}_m)\sum_{\kappa'=1}^{c_m} \sum_{i \in C_m} u^{(m)}_{\kappa i} u^{(m)}_{\kappa' i}{\boldsymbol \eta}^{(m)}_{\kappa'}
- \sigma \sum_{m'=1}^M \widetilde{A}_{mm'}D{\bold H}({\bold s}_{m'}-{\bold s}_m) \sum_{\kappa'=1}^{c_m} \sum_{i \in C_m} u^{(m)}_{\kappa i} u^{(m)}_{\kappa' i}{\boldsymbol \eta}^{(m)}_{\kappa'} \\
&\quad\quad+ \sigma\sum_{m'=1}^MD{\bold H}({\bold s}_{m'}-{\bold s}_m)\sum_{\kappa'=1}^{c_{m'}}\sum_{i \in C_m}\sum_{j \in C_{m'}}u^{(m)}_{\kappa i}A_{ij}u^{(m')}_{\kappa'j}{\boldsymbol \eta}^{(m')}_{\kappa'} \\
&= D{\bold F}({\bold s}_m){\boldsymbol \eta}^{(m)}_{\kappa}-\sigma\sum_{m'=1}^M \widetilde{A}_{mm'}D{\bold H}({\bold s}_{m'}-{\bold s}_m){\boldsymbol \eta}^{(m)}_{\kappa}+
\sigma\sum_{m'=1}^{M}D{\bold H}({\bold s}_{m'}-{\bold s}_m)\sum_{\kappa'=1}^{c_{m'}}B^{(mm')}_{\kappa\kappa'}{\boldsymbol \eta}^{(m')}_{\kappa'} \\
&= \Big[D{\bold F}({\bold s}_m)-\sigma\sum_{m'=1}^M \widetilde{A}_{mm'}D{\bold H}({\bold s}_{m'}-{\bold s}_m)\Big]{\boldsymbol \eta}^{(m)}_{\kappa}+
\sigma\sum_{m'=1}^{M'}\sum_{\kappa'=2}^{c_{m'}} B^{(mm')}_{\kappa\kappa'} D{\bold H}({\bold s}_{m'}-{\bold s}_m){\boldsymbol \eta}^{(m')}_{\kappa'}
\end{split}
\label{diffusive_variation_block}
\end{equation}
for $m=1,\ldots,M$ and $\kappa = 2,\ldots,c_m$.
When the basis vectors $\bold{u}^{(m)}_{\kappa}$ with $\kappa \ge 2$ corresponding to ISCs are chosen to be eigenvectors of $A$, as done in Sec.~\ref{sec:Eq(3-4)}, the equation becomes
\begin{equation}
\begin{split}
\dot{\boldsymbol \eta}^{(m)}_{\kappa}
&= \Big[D{\bold F}({\bold s}_m)-\sigma\sum_{m'=1}^M \widetilde{A}_{mm'}D{\bold H}({\bold s}_{m'}-{\bold s}_m)\Big]{\boldsymbol \eta}^{(m)}_{\kappa}+
\sigma\sum_{\kappa'=2}^{c_{m}} B^{(mm)}_{\kappa\kappa'} D{\bold H}({\bold 0}){\boldsymbol \eta}^{(m)}_{\kappa'} \\ 
&= \Big[D{\bold F}({\bold s}_m)-\sigma\sum_{m'=1}^M \widetilde{A}_{mm'}D{\bold H}({\bold s}_{m'}-{\bold s}_m)+
\sigma D{\bold H}({\bold 0})\lambda^{(m)}_{\kappa}\Big]{\boldsymbol \eta}^{(m)}_{\kappa}. 
\end{split}
\label{diffusive_variation_eigenvector}
\end{equation}
In this case, the stability of an ISC or ISC set explicitly depends on the states ${\bold s}_{m'}$ of the clusters in other ISCs or ISC sets due to the presence of the second term in the square brackets, which stems from the diffusive nature of the coupling in Eq.~\eqref{diffusive_rate}.

\section{A\lowercase{pplication:} P\lowercase{ermanently stable chimeras in electro-optic networks}}

\subsection{State and stability equations}

We consider the electro-optic system described by Eq.~(5) of the main text.
For this system, the equation for a cluster synchronous state [Eq.~(2)], reads
\begin{equation}\label{eqn:EO_reduced}
s_m(t+1) = \biggl[ \beta \mathcal{I}(s_m(t)) + \sigma \sum_{m'=1}^M \widetilde{A}_{mm'} \mathcal{I}(s_{m'}(t)) + \delta \biggr] \,\text{mod $2\pi$},
\end{equation}
where we recall $\mathcal{I}(x)=[1-{\mathrm {cos}}(x)]/2$.
The decoupled stability equation [Eq.~(3)] reads
\begin{equation}
\begin{split}
\eta^{(m)}_{\kappa}(t+1) 
&= \beta \, \mathcal{I}'(s_m(t)) \cdot \eta_{\kappa}^{(m)}(t) +
\sigma\sum_{m'=1}^{M'}\sum_{\kappa'=2}^{c_{m'}} \mathcal{I}'(s_{m'}(t)) B^{(mm')}_{\kappa\kappa'} \cdot \eta_{\kappa'}^{(m')}(t) \\
&= \frac{\beta}{2} \, \sin(s_m(t)) \cdot \eta_{\kappa}^{(m)}(t) +
\frac{\sigma}{2} \sum_{m'=1}^{M'}\sum_{\kappa'=2}^{c_{m'}} B^{(mm')}_{\kappa\kappa'} \sin(s_{m'}(t)) \cdot \eta_{\kappa'}^{(m')}(t),
\label{Eq:pecora_variation_block}
\end{split}
\end{equation}
while the diagonalized version of the equation for ISCs [Eq.~(4)] reads
\begin{equation}\label{eqn:stab_eq}
\begin{split}
\eta_{\kappa}^{(m)}(t+1) 
&= \Bigl[ \beta \, \mathcal{I}'(s_m(t)) + \sigma\lambda_{\kappa}^{(m)} \, \mathcal{I}'(s_m(t)) \Bigr] \cdot \eta_{\kappa}^{(m)}(t) \\
&= \frac{1}{2} \bigl( \beta + \sigma\lambda_{\kappa}^{(m)} \bigr) \sin(s_m(t)) \cdot \eta_{\kappa}^{(m)}(t).
\end{split}
\end{equation}

\begin{figure}[t]
\includegraphics[width=0.8\linewidth]{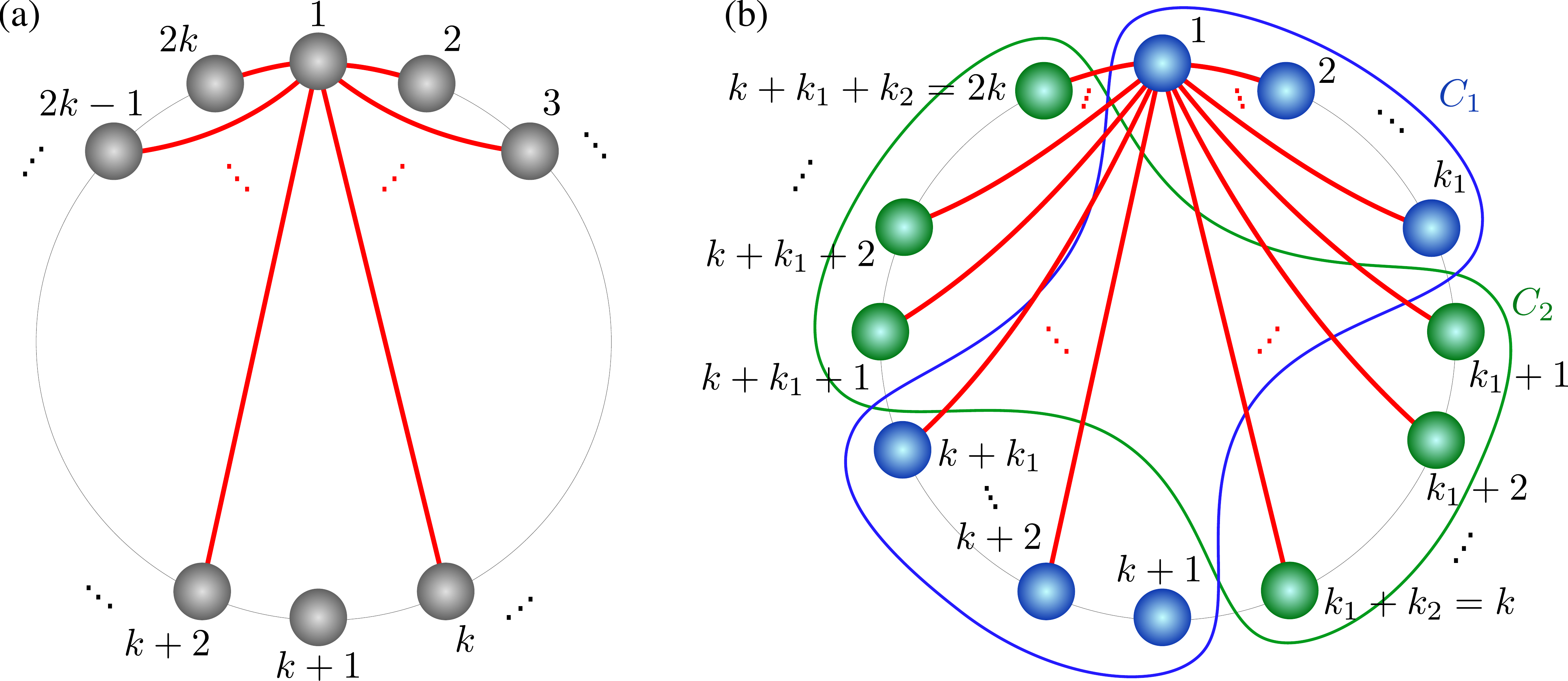}
\caption{Ring network of $N=2k$ nodes, each with $2(k-1)$ neighbors.
(a) Network structure.
Node $1$ is connected to all the other nodes except for node $k+1$, located on the opposite side of the ring (red links).
Each node is connected to the rest of the network in the same manner (not shown for visual clarity).
(b) Class of two-cluster candidate CS patterns for the network, in which clusters $C_1$ and $C_2$ have $2k_1$ and $2k_2$ nodes, respectively (where $k_1+k_2=k$).}
\label{Fig:fig_ring_network}
\end{figure}

For the network structure, we consider a ring of $N=2k$ nodes, each connected to $2(k-1)$ neighbors, as illustrated in Fig.~\ref{Fig:fig_ring_network}(a).

The $N \times N$ adjacency matrix is given by\\[0pt]
\begin{equation}\label{eqn:A}
A=
\begin{bmatrix}
 \makebox[0pt][l]{$\smash{\overbrace{\phantom{%
   \begin{matrix}\phantom{\rule{55pt}{0pt}}\end{matrix}}}^{k}}$}
 0      & 1      & \cdots & 1      & \vline & 
 \makebox[0pt][l]{$\smash{\overbrace{\phantom{%
   \begin{matrix}\phantom{\rule{53pt}{0pt}}\end{matrix}}}^{k}}$}
 0      & 1      & \cdots & 1       \\[-5pt]
 1      & 0      & \ddots & \vdots & \vline & 1      & 0      & \ddots & \vdots \\[-5pt]
 \vdots & \ddots & \ddots & 1      & \vline & \vdots & \ddots & \ddots & 1      \\
 1      & \cdots & 1      & 0      & \vline & 1      & \cdots & 1      & 0      \\
 \hline
 0      & 1      & \cdots & 1      & \vline & 0      & 1      & \cdots & 1      \\[-5pt]
 1      & 0      & \ddots & \vdots & \vline & 1      & 0      & \ddots & \vdots \\[-5pt]
 \vdots & \ddots & \ddots & 1      & \vline & \vdots & \ddots & \ddots & 1      \\
 1      & \cdots & 1      & 0      & \vline & 1      & \cdots & 1      & 0      \\
\end{bmatrix}.
\end{equation}
This network has a number of candidate CS patterns, among which is the one with a single cluster $C_1 = \{ 1,\ldots,2k \}$ (corresponding to the complete synchronization of all nodes) and a class of two-cluster patterns with
\begin{equation}\label{eqn:two_cluster_pattern}
\begin{split}
\phantom{\rule{0pt}{20pt}}
C_1 &= \{ 
\makebox[0pt][l]{$\smash{\overbrace{\phantom{\begin{matrix}\phantom{\rule{30pt}{0pt}}\end{matrix}}}^{k_1}}$}
1,\ldots,k_1, 
\,\,\makebox[0pt][l]{$\smash{\overbrace{\phantom{\begin{matrix}\phantom{\rule{70pt}{0pt}}\end{matrix}}}^{k_1}}$}
k+1,\ldots,k+k_1 \},\\
C_2 &= \{ 
\makebox[0pt][l]{$\smash{\underbrace{\phantom{\begin{matrix}\phantom{\rule{54pt}{0pt}}\end{matrix}}}_{k_2}}$}
k_1+1,\ldots,k, 
\,\,\makebox[0pt][l]{$\smash{\underbrace{\phantom{\begin{matrix}\phantom{\rule{77pt}{0pt}}\end{matrix}}}_{k_2}}$}
k+k_1+1,\ldots,2k \},\\[10pt]
\end{split}
\end{equation}
which are illustrated in Fig.~\ref{Fig:fig_ring_network}(b).

\subsection{Stability of complete synchronization}

The completely synchronous state, $x_i(t)=s(t)$, $i=1,\ldots,N$, can be regarded as a special case of cluster synchronization with CS pattern consisting of a single cluster $C_1 = \{ 1,\ldots,N \}$ (which is an ISC).
Thus, $s(t)$ satisfies Eq.~\eqref{eqn:EO_reduced}, which for this CS pattern reads
\begin{equation}\label{eqn:EO_reduced_comp_sync}
s(t+1) = \Bigl[ \bigl( \beta + \sigma(N-2) \bigr) \, \mathcal{I}(s(t)) + \delta\Bigr] \,\text{mod $2\pi$}.
\end{equation}
To analyze the stability of this state, we choose the columns of the coordinate transformation matrix $U$ in Eq.~\eqref{eqn:U} to be eigenvectors of $A$:
\begin{equation}\label{eqn:EO_comp_sync_U}
U = \begin{bmatrix}
1/\sqrt{N} & \vline & u_{1,2} & \cdots & u_{1,k} & \vline & 1/\sqrt{N} & u_{1,2} & \cdots & u_{1,N}\\
\vdots & \vline & \vdots & & \vdots & \vline & \vdots & \vdots & & \vdots\\
1/\sqrt{N} & \vline & u_{k,2} & \cdots & u_{k,k} & \vline & 1/\sqrt{N} & u_{k,2} & \cdots & u_{k,N}\\
\hline
1/\sqrt{N} & \vline & u_{1,2} & \cdots & u_{1,k} & \vline & -1/\sqrt{N} & -u_{1,2} & \cdots & -u_{1,N}\\
\vdots & \vline & \vdots & & \vdots & \vline & \vdots & \vdots & & \vdots\\
1/\sqrt{N} & \vline & u_{k,2} & \cdots & u_{k,k} & \vline & -1/\sqrt{N} & -u_{k,2} & \cdots & -u_{k,N}\\
\end{bmatrix},
\end{equation}
where $u_{i,\kappa}$ are selected to satisfy
\begin{equation}
u_{1,\kappa} + \cdots + u_{k,\kappa} = 0,
\quad
u_{1,\kappa}^2 + \cdots + u_{k,\kappa}^2 = 1,
\quad
u_{1,\kappa}u_{1,\kappa'} + \cdots + u_{k,\kappa}u_{k,\kappa'} = 0
\end{equation}
for each $\kappa,\kappa' = 2,\ldots,k$, $\kappa\neq\kappa'$, to ensure the orthonormality of the column vectors.
This $U$ fully diagonalize $A$, with the eigenvalues $N-2$, $-2$, and $0$ appearing on the diagonal:\\[0pt]
\begin{equation}
B=UAU^{-1}=
\left[\,\,\begin{array}{|c|ccc|ccc|}
 \cline{1-1}
 N\!\!-\!2 \,\rule{0pt}{12pt} & \\[3pt]
 \cline{1-4}
 \multicolumn{1}{c|}{} & -2 & & \\[-5pt]
 \multicolumn{1}{c|}{} & & \ddots & \\[-5pt]
 \multicolumn{1}{c|}{} & & & \!\!-2 \\
 \cline{2-7}
 \multicolumn{4}{c|}{} & 0 & & \\[-5pt]
 \multicolumn{4}{c|}{} & & \ddots & \\[-5pt]
 \multicolumn{4}{c|}{} & & & 0 \\
 \cline{5-7}
\end{array}\,\,\right].
\end{equation}
With the matrix $U$ in Eq.~\eqref{eqn:EO_comp_sync_U}, the stability equation~\eqref{eqn:stab_eq} becomes
\begin{align}
\eta_{\kappa}(t+1) &= \Bigl(\frac{\beta}{2}-\sigma\Bigr) \sin(s(t)) \cdot \eta_{\kappa}(t), & &\kappa=2,\ldots,k, \label{eqn:stabilty_comp_sync_1} \\
\eta_{\kappa}(t+1) &= \frac{\beta}{2} \sin(s(t)) \cdot \eta_{\kappa}(t), & &\kappa=k+1,\ldots,2k, \label{eqn:stabilty_comp_sync_2}
\end{align}
where we suppressed the cluster index $m$ because there is only one cluster in this case.

Using these results, we now derive a sufficient condition that implies condition~(i) for permanent chimera states in the main text (i.e., that the complete synchronization of the network is unstable).
On the one hand, from Eq.~\eqref{eqn:EO_reduced_comp_sync} and the fact that $0 \le \mathcal{I}(x)=[1-{\mathrm {cos}}(x)]/2 \le 1$, we see that $s(t)$ remains in the interval $I := [\delta, \, \beta+\sigma(N-2)+\delta]$ if $\beta + \sigma(N-2) > 0$.
On the other hand, from Eqs.~\eqref{eqn:stabilty_comp_sync_1} and \eqref{eqn:stabilty_comp_sync_2}, the complete synchronization is unstable if either $\bigl|\bigl(\frac{\beta}{2}-\sigma\bigr)\sin(s(t))\bigr| > 1$ for all $t$ or $\frac{|\beta|}{2}|\sin(s(t))| > 1$ for all $t$.
Combining these two assertions, a sufficient condition can be expressed as
\begin{equation}
\min_{x \in I} \, \bigl|\sin(x)\bigr| > 2 \cdot \min\!\left\{ \frac{1}{|\beta - 2\sigma|}, \frac{1}{|\beta|} \right\}.
\end{equation}
For $\beta + \sigma(N-2) = \frac{2\pi}{3} > 0$ and $\delta = \frac{\pi}{6}$, we have $I = [\frac{\pi}{6}, \, \frac{5\pi}{6}]$, and hence $\min_{x \in I} \, \bigl|\sin(x)\bigr| = \frac{1}{2}$.
In this case, the sufficient condition above can be shown to be satisfied if $\sigma < - \frac{6-\pi}{3k}$ or $\sigma > \frac{6+\pi}{3k}$.
Thus, a set of parameter choices that guarantee that the complete synchronization is unstable can be expressed as
\begin{equation}\label{eqn:candidate_param}
\sigma \in \Bigl(-\infty, -\frac{6-\pi}{3k} \Bigr) \cup \Bigl( \frac{6+\pi}{3k}, \infty \Bigr), \quad
\beta = \frac{2\pi}{3} - 2\sigma(k-1), \quad
\delta = \frac{\pi}{6}.
\end{equation}
One can search in this set for cases that also satisfy condition~(ii) for permanent chimera states in the main text, which would require analyzing the synchronization stability of clusters in other CS patterns.

\subsection{Stability of two-cluster CS patterns}
\label{sec:two_cluster_stability}

For the candidate CS pattern of the form Eq.~\eqref{eqn:two_cluster_pattern} [illustrated in Fig.~\ref{Fig:fig_ring_network}(b)], we have
\begin{equation}
\widetilde{A} = \begin{bmatrix}
2(k_1 - 1) & 2k_2 \\
2k_1     & 2(k_2 - 1)
\end{bmatrix}.
\end{equation}
With this, Eq.~\eqref{eqn:EO_reduced} for the synchronous state, $x_i(t) = s_m(t)$, $i \in C_m$, becomes 
\begin{equation}\label{eqn:EO_reduced_two_cluster}
\begin{split}
s_1(t+1) &= \Bigl[ \bigl( \beta + 2\sigma(k_1 - 1) \bigl) \mathcal{I}(s_1(t)) + 2\sigma k_2 \mathcal{I}(s_2(t)) + \delta \Bigr] \,\text{mod $2\pi$,} \\
s_2(t+1) &= \Bigl[ \bigl( \beta + 2\sigma(k_2 - 1) \bigl) \mathcal{I}(s_2(t)) + 2\sigma k_1 \mathcal{I}(s_1(t)) + \delta \Bigr] \,\text{mod $2\pi$.}
\end{split}
\end{equation}
We note that the adjacency matrix in Eq.~\eqref{eqn:A} has the following special structure associated with the two-cluster pattern:
\begin{equation}
A=
\begin{bmatrix}
 \makebox[0pt][l]{$\smash{\overbrace{\phantom{%
   \begin{matrix}\phantom{\rule{55pt}{0pt}}\end{matrix}}}^{k_1}}$}
 0      & 1      & \cdots & 1      & \vline & 
 \makebox[0pt][l]{$\smash{\overbrace{\phantom{%
   \begin{matrix}\phantom{\rule{53pt}{0pt}}\end{matrix}}}^{k_2}}$}
 1      & \cdots & \cdots & 1      & \vline &
 \makebox[0pt][l]{$\smash{\overbrace{\phantom{%
   \begin{matrix}\phantom{\rule{55pt}{0pt}}\end{matrix}}}^{k_1}}$}
 0      & 1      & \cdots & 1      & \vline & 
 \makebox[0pt][l]{$\smash{\overbrace{\phantom{%
   \begin{matrix}\phantom{\rule{53pt}{0pt}}\end{matrix}}}^{k_2}}$}
 1      & \cdots & \cdots & 1 \\[-5pt]
 1      & 0      & \ddots & \vdots & \vline & \vdots &        &        & \vdots & \vline &
 1      & 0      & \ddots & \vdots & \vline & \vdots &        &        & \vdots \\[-5pt]
 \vdots & \ddots & \ddots & 1      & \vline & \vdots &        &        & \vdots & \vline &
 \vdots & \ddots & \ddots & 1      & \vline & \vdots &        &        & \vdots \\[-5pt]
 1      & \cdots & 1      & 0      & \vline & 1      & \cdots & \cdots & 1      & \vline &
 1      & \cdots & 1      & 0      & \vline & 1      & \cdots & \cdots & 1      \\
 \hline
 1      & \cdots & \cdots & 1      & \vline & 0      & 1      & \cdots & 1      & \vline &
 1      & \cdots & \cdots & 1      & \vline & 0      & 1      & \cdots & 1 \\[-5pt]
 \vdots &        &        & \vdots & \vline & 1      & 0      & \ddots & \vdots & \vline &
 \vdots &        &        & \vdots & \vline & 1      & 0      & \ddots & \vdots \\[-5pt]
 \vdots &        &        & \vdots & \vline & \vdots & \ddots & \ddots & 1      & \vline &
 \vdots &        &        & \vdots & \vline & \vdots & \ddots & \ddots & 1      \\[-5pt]
 1      & \cdots & \cdots & 1      & \vline & 1      & \cdots & 1      & 0      & \vline &
 1      & \cdots & \cdots & 1      & \vline & 1      & \cdots & 1      & 0      \\
 \hline
 0      & 1      & \cdots & 1      & \vline & 1      & \cdots & \cdots & 1      & \vline &
 0      & 1      & \cdots & 1      & \vline & 1      & \cdots & \cdots & 1 \\[-5pt]
 1      & 0      & \ddots & \vdots & \vline & \vdots &        &        & \vdots & \vline &
 1      & 0      & \ddots & \vdots & \vline & \vdots &        &        & \vdots \\[-5pt]
 \vdots & \ddots & \ddots & 1      & \vline & \vdots &        &        & \vdots & \vline &
 \vdots & \ddots & \ddots & 1      & \vline & \vdots &        &        & \vdots \\[-5pt]
 1      & \cdots & 1      & 0      & \vline & 1      & \cdots & \cdots & 1      & \vline &
 1      & \cdots & 1      & 0      & \vline & 1      & \cdots & \cdots & 1      \\
 \hline
 1      & \cdots & \cdots & 1      & \vline & 0      & 1      & \cdots & 1      & \vline &
 1      & \cdots & \cdots & 1      & \vline & 0      & 1      & \cdots & 1 \\[-5pt]
 \vdots &        &        & \vdots & \vline & 1      & 0      & \ddots & \vdots & \vline &
 \vdots &        &        & \vdots & \vline & 1      & 0      & \ddots & \vdots \\[-5pt]
 \vdots &        &        & \vdots & \vline & \vdots & \ddots & \ddots & 1      & \vline &
 \vdots &        &        & \vdots & \vline & \vdots & \ddots & \ddots & 1      \\[-5pt]
 1      & \cdots & \cdots & 1      & \vline & 1      & \cdots & 1      & 0      & \vline &
 1      & \cdots & \cdots & 1      & \vline & 1      & \cdots & 1      & 0      \\
 \end{bmatrix}.
\end{equation}
Taking advantage of this form, the cluster-based coordinate transformation matrix $U$ can be chosen as
\begin{equation}\label{eqn:EO_two_cluster_U}
\begin{split}
U^T 
&= \Bigl[{\bold{u}^{(1)}_{1}},{\bold{u}^{(2)}_{1}}\,\Big\vert\,
{\bold{u}^{(1)}_{2}},\ldots,{\bold{u}^{(1)}_{k_1}}\,\Big\vert\,{\bold{u}^{(1)}_{k_1+1}},\ldots,{\bold{u}^{(1)}_{2k_1}}\,\Big\vert\,
{\bold{u}^{(2)}_{2}},\ldots,{\bold{u}^{(2)}_{k_2}}\,\Big\vert\,{\bold{u}^{(2)}_{k_2+1}},\ldots,{\bold{u}^{(2)}_{2k_2}} \Bigr]\\[5pt]
&= \begin{bmatrix}
\frac{1}{\sqrt{2k_1}} & & \vline & u_{1,2}^{(1)} & \cdots & u_{1,k_1}^{(1)} & \vline 
 & \frac{1}{\sqrt{2k_1}} & u_{1,2}^{(1)} & \cdots & u_{1,k_1}^{(1)} & \vline & & & & \vline \\
\vdots & & \vline & \vdots & & \vdots & \vline & \vdots & \vdots & & \vdots & \vline & & & & \vline\\
\frac{1}{\sqrt{2k_1}} & & \vline & u_{k_1,2}^{(1)} & \cdots & u_{k_1,k_1}^{(1)} & \vline 
 & \frac{1}{\sqrt{2k_1}} & u_{k_1,2}^{(1)} & \cdots & u_{k_1,k_1}^{(1)}
 & \vline & & & & \vline \\[5pt]
\hline
 \rule{0pt}{12pt} & \frac{1}{\sqrt{2k_2}} & \vline & & & & \vline & & & & 
 & \vline & u_{1,2}^{(2)} & \cdots & u_{1,k_2}^{(2)} & \vline & \frac{1}{\sqrt{2k_2}} & u_{1,2}^{(2)} & \cdots & u_{1,k_2}^{(2)} \\
   & \vdots & \vline & & & & \vline & & & & & \vline 
 & \vdots & & \vdots & \vline & \vdots & \vdots & & \vdots\\
   & \frac{1}{\sqrt{2k_2}} & \vline & & & & \vline 
 & & & & & \vline & u_{k_2,2}^{(2)} & \cdots & u_{k_2,k_2}^{(2)} & \vline & \frac{1}{\sqrt{2k_2}} & u_{k_2,2}^{(2)} & \cdots & u_{k_2,k_2}^{(2)} \\[5pt]
\hline
\rule{0pt}{12pt} \frac{1}{\sqrt{2k_1}} & & \vline & u_{1,2}^{(1)} & \cdots & u_{1,k_1}^{(1)} & \vline 
 & \frac{-1}{\sqrt{2k_1}} & -u_{1,2}^{(1)} & \cdots & -u_{1,k_1}^{(1)} & \vline & & & & \vline \\
\vdots & & \vline & \vdots & & \vdots & \vline & \vdots & \vdots & & \vdots & \vline & & & & \vline\\
\frac{1}{\sqrt{2k_1}} & & \vline & u_{k_1,2}^{(1)} & \cdots & u_{k_1,k_1}^{(1)} & \vline 
 & \frac{-1}{\sqrt{2k_1}} & -u_{k_1,2}^{(1)} & \cdots & -u_{k_1,k_1}^{(1)}
 & \vline & & & & \vline \\[5pt]
\hline
 \rule{0pt}{12pt}  & \frac{1}{\sqrt{2k_2}} & \vline & & & & \vline & & & & 
 & \vline & u_{1,2}^{(2)} & \cdots & u_{1,k_2}^{(2)} & \vline & \frac{-1}{\sqrt{2k_2}} & -u_{1,2}^{(2)} & \cdots & -u_{1,k_2}^{(2)} \\
   & \vdots & \vline & & & & \vline & & & & & \vline 
 & \vdots & & \vdots & \vline & \vdots & \vdots & & \vdots\\
   & \frac{1}{\sqrt{2k_2}} & \vline & & & & \vline 
 & & & & & \vline & u_{k_2,2}^{(2)} & \cdots & u_{k_2,k_2}^{(2)} & \vline & \frac{-1}{\sqrt{2k_2}} & -u_{k_2,2}^{(2)} & \cdots & -u_{k_2,k_2}^{(2)} \\
\end{bmatrix},
\end{split}
\end{equation}
where $u_{i,\kappa}^{(m)}$ are selected to satisfy
\begin{equation}\label{eqn:EO_three_cluster_u_eqn_1}
u_{1,\kappa}^{(m)} + \cdots + u_{k_m,\kappa}^{(m)} = 0,
\quad
\bigl(u_{1,\kappa}^{(m)}\bigr)^2 + \cdots + \bigl(u_{k_m,\kappa}^{(m)}\bigr)^2 = 1,
\quad
u_{1,\kappa}^{(m)} u_{1,\kappa'}^{(m)} + \cdots + u_{k_m,\kappa}^{(m)} u_{k_m,\kappa'}^{(m)} = 0
\end{equation}
for each $\kappa,\kappa' = 2,\ldots,k_m$, $\kappa\neq\kappa'$, $m=1,2$, to ensure the orthonormality of the column vectors.
This choice of basis vectors guarantees that $\bold{u}^{(m)}_2,\ldots,\bold{u}^{(m)}_{k_m}$ and $\bold{u}^{(m)}_{k_m+1},\ldots,\bold{u}^{(m)}_{2k_m}$ are eigenvectors of $A$ associated with eigenvalues $\lambda^{(m)}_{\kappa} = -2$ and $\lambda^{(m)}_{\kappa} = 0$, respectively. 
With this coordinate transformation matrix $U$, the adjacency matrix $A$ is block diagonalized as
\begin{equation}
B=UAU^{-1}=
\left[\,\,\begin{array}{|cc|cccccc|cccccc|}
 \cline{1-2}
 2(k_1-1) & 2\sqrt{k_1 k_2} \rule{0pt}{12pt} \\[2pt]
 2\sqrt{k_1 k_2} & 2(k_2-1) \\[3pt]
 \cline{1-8}
 \multicolumn{2}{c|}{} & 
 \makebox[0pt][l]{$\smash{\overbrace{\phantom{%
  \begin{matrix}\phantom{\rule{47pt}{14pt}}\end{matrix}}}^{k_1-1}}$}
 \!\!-2 & & & & &\\[-5pt]
 \multicolumn{2}{c|}{} & & \ddots & & & &\\[-5pt]
 \multicolumn{2}{c|}{} & & & \!\!\!-2 & & & \\[-5pt]
 \multicolumn{2}{c|}{} & & & & 
 \makebox[0pt][l]{$\smash{\overbrace{\phantom{%
  \begin{matrix}\phantom{\rule{35pt}{0pt}}\end{matrix}}}^{k_1}}$}
 0 & & \\[-5pt]
 \multicolumn{2}{c|}{} & & & & & \ddots & \\[-5pt]
 \multicolumn{2}{c|}{} & & & & & & 0 \\
 \cline{3-14}
 \multicolumn{8}{c|}{} & 
 \makebox[0pt][l]{$\smash{\overbrace{\phantom{%
  \begin{matrix}\phantom{\rule{47pt}{14pt}}\end{matrix}}}^{k_2-1}}$}
 \!\!-2 & & & & &\\[-5pt]
 \multicolumn{8}{c|}{} & & \ddots & & & &\\[-5pt]
 \multicolumn{8}{c|}{} & & & \!\!\!-2 & & & \\[-5pt]
 \multicolumn{8}{c|}{} & & & & 
 \makebox[0pt][l]{$\smash{\overbrace{\phantom{%
  \begin{matrix}\phantom{\rule{35pt}{0pt}}\end{matrix}}}^{k_2}}$}
 0 & & \\[-5pt]
 \multicolumn{8}{c|}{} & & & & & \ddots & \\[-5pt]
 \multicolumn{8}{c|}{} & & & & & & 0 \\
 \cline{9-14}
\end{array}\,\,\right],
\end{equation}
and the stability equation~\eqref{eqn:stab_eq} becomes
\begin{align}
\eta_{\kappa}^{(1)}(t+1) 
&= \Bigl(\frac{\beta}{2}-\sigma\Bigr) \sin(s_1(t)) \cdot \eta_{\kappa}^{(1)}(t), & &\kappa=2,\ldots,k_1, \label{eqn:stabilty_2c_1}\\
\eta_{\kappa}^{(1)}(t+1) &= \frac{\beta}{2} \sin(s_1(t)) \cdot \eta_{\kappa}^{(1)}(t), & &\kappa=k_1+1,\ldots,2k_1, \label{eqn:stabilty_2c_2}\\
\eta_{\kappa}^{(2)}(t+1)
&= \Bigl(\frac{\beta}{2}-\sigma\Bigr) \sin(s_2(t)) \cdot \eta_{\kappa}^{(2)}(t), & &\kappa=2,\ldots,k_2, \label{eqn:stabilty_2c_3}\\
\eta_{\kappa}^{(2)}(t+1) &= \frac{\beta}{2} \sin(s_2(t)) \cdot \eta_{\kappa}^{(2)}(t), & &\kappa=k_2+1,\ldots,2k_2. \label{eqn:stabilty_2c_4}
\end{align}
These equations, together with $s_1(t)$ and $s_2(t)$ computed by iterating Eq.~\eqref{eqn:EO_reduced_two_cluster}, define the Lyapunov exponents $\Lambda_{\kappa}^{(m)}$ that determine the transverse stability of the two-cluster synchronous state.
There are a total of $(N-2)$ transverse Lyapunov exponents, but since there are only four types of stability equations, we have at most four distinct values for these exponents:
\begin{equation}
\Lambda_{2}^{(1)} = \cdots = \Lambda_{k_1}^{(1)}, \quad
\Lambda_{k_1+1}^{(1)} = \cdots = \Lambda_{2k_1}^{(1)}, \quad
\Lambda_{2}^{(2)} = \cdots = \Lambda_{k_2}^{(2)}, \quad
\Lambda_{k_2+1}^{(2)} = \cdots = \Lambda_{2k_2}^{(2)}.
\end{equation}
The exponents $\Lambda_{2}^{(1)}$ and $\Lambda_{k_1+1}^{(1)}$ determines the synchronization stability of cluster $C_1$, which is independent from the synchronization stability of $C_2$, determined by $\Lambda_{2}^{(2)}$ and $\Lambda_{k_2+1}^{(2)}$.

\subsection{Example: The chimera state of Fig.~3}

\begin{figure}[t]
\includegraphics[width=0.75\linewidth]{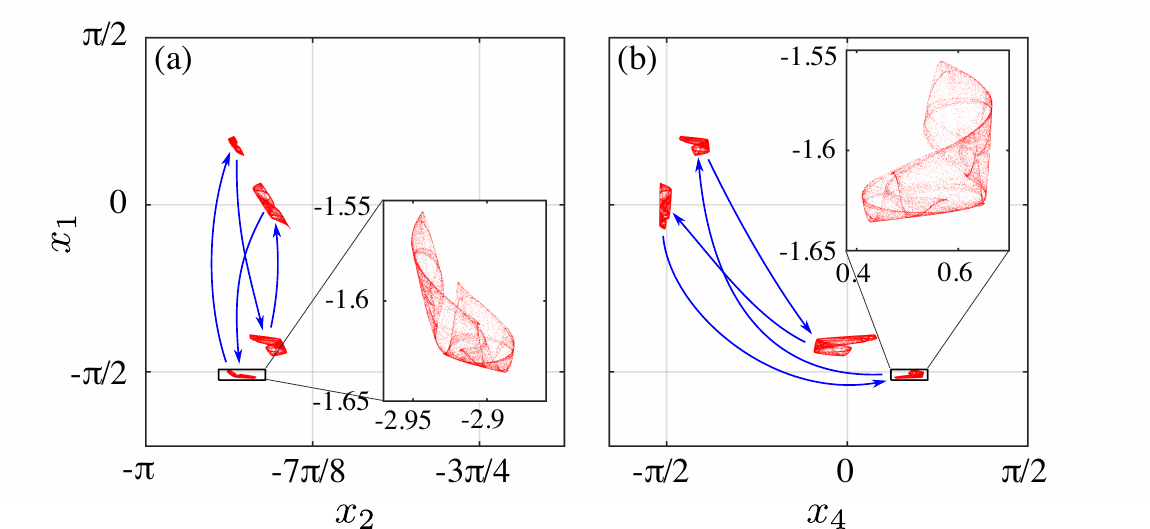}
\caption{Four-band chaotic dynamics of the system considered in Fig.~3 of the main text.
The state variable $x_1$ is plotted against $x_2$ (a) and $x_4$ (b) using the last $10^5$ iterations of the same trajectory (of length $10^9$) used in Fig.~3(c--d).
To show the continuity of the chaotic bands in the full phase space (which is periodic, with $x_i=0$ and $x_i=2\pi$ identified), we translate each $x_i$ as $x_i \to x_i-2\pi$ when $\pi < x_i \le 2\pi$ to change the numerical range of $x_i$ to $[-\pi,\pi]$.
The blue arrows indicate the order in which the trajectory visits the four bands.}
\label{Fig:Chimera_EO_regular_N_6_k_4_sm}
\end{figure}

In this case the network is of size $N=2k=6$, and the candidate set of parameters in Eq.~\eqref{eqn:candidate_param}, which ensures that complete synchronization is unstable, are given by $\beta = \frac{2\pi}{3} - 4\sigma$, $\delta = \frac{\pi}{6}$, and either $\sigma < -\frac{6-\pi}{9} \approx -0.32$ or $\sigma > \frac{6+\pi}{9} \approx 1.02$.
The CS pattern considered in Fig.~3 (with two ISCs, $C_1 = \{1,4\}$ and $C_2 = \{2,3,5,6\}$, associated with the subgroups $G_1 = \langle(1,4)\rangle$ and $G_2 = \langle(2,6)(3,5),(2,3)(6,5)\rangle$, respectively)
is a special case of the two-cluster patterns analyzed in the previous section, in which $k_1 = 1$ and $k_2 = 2$.
The transverse Lyapunov exponents $\Lambda_{2}^{(1)}$, $\Lambda_{2}^{(2)}$, and $\Lambda_{3}^{(2)}$ ($= \Lambda_{4}^{(2)}$), shown in Fig.~3(b) for a range of $\sigma$, are numerically determined from Eq.~\eqref{eqn:EO_reduced_two_cluster} and Eqs.~(\ref{eqn:stabilty_2c_1}--\ref{eqn:stabilty_2c_4}).
We take $\sigma = -0.55$ [indicated by the vertical dashed line in Fig.~3(b)], which makes the synchronization stable for $C_2$ (i.e., $\Lambda^{(2)}_2 < 0$ and $\Lambda^{(2)}_3 = \Lambda^{(2)}_4 < 0$) and unstable for $C_1$ (i.e., $\Lambda^{(1)}_2 > 0$).
Since $C_1$ (being of size two) cannot be broken into smaller nontrivial clusters, condition~(ii) for permanent chimera states is satisfied for this parameter choice.
A trajectory starting from a random initial condition, shown in Fig.~3(a), indeed exhibits chimera behavior up to $t = 10^9$.
We observe that the system converges to a four-band chaotic attractor within the synchronization manifold of the cluster $C_2$ given by $\{ (x_1,\dots,x_6) \,|\, x_2=x_3=x_5=x_6 \}$, which is shown in Fig.~\ref{Fig:Chimera_EO_regular_N_6_k_4_sm}.
This in particular eliminates the possibility that the cluster $C_1$ is in a state of generalized synchronization, in which $x_1$ and $x_4$ are related by a function~\cite{sync_book}.
Since the conditions we discussed above are satisfied for a range of $\sigma$, $\beta$ and $\delta$, we expect permanently stable chimera states to be common in this system and be robust against parameter changes.

\subsection{Example: Chimera state in a larger network}
\label{sec:chimera_ring_N_10_k_8_N1_2}

\begin{figure}[h]
\includegraphics[width=0.45\linewidth]{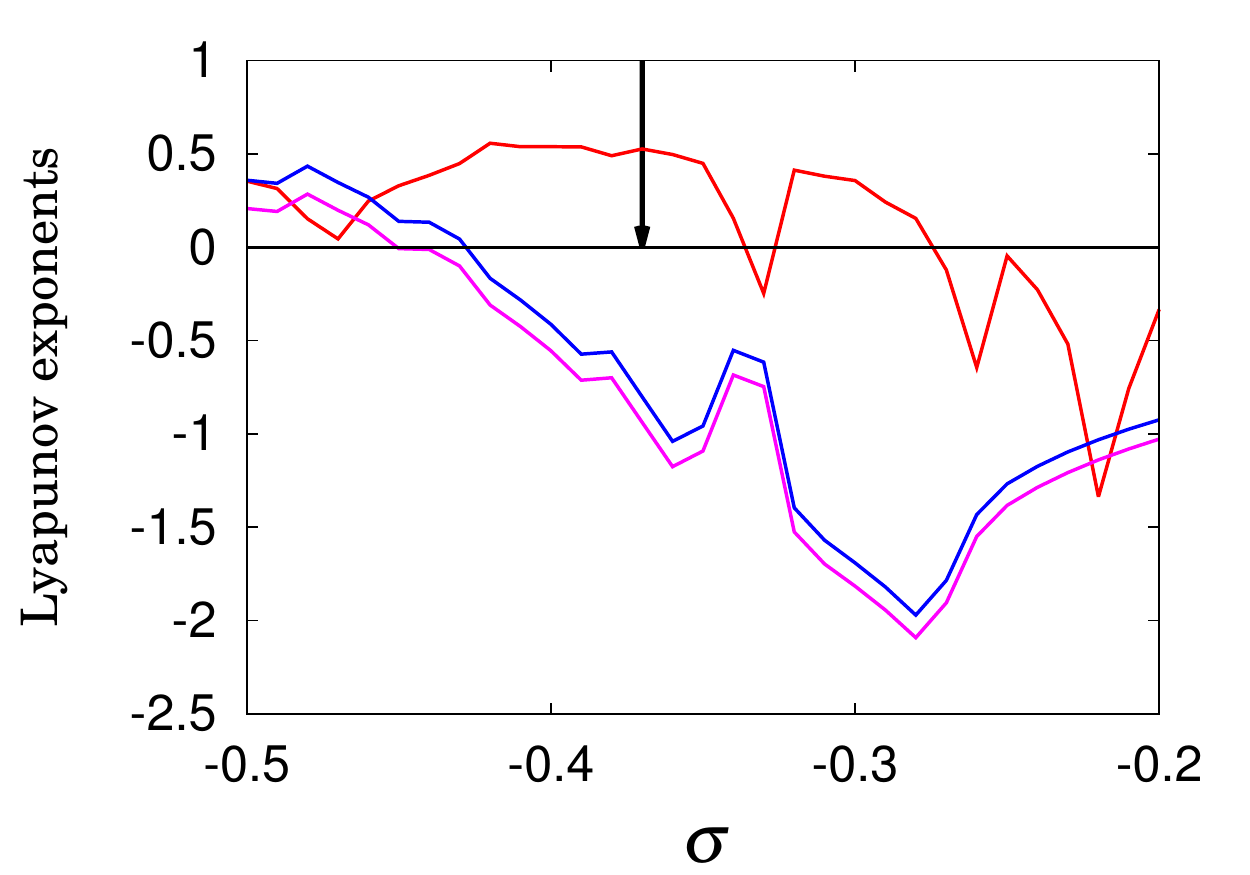}
\vspace{-8pt}
\caption{Transverse Lyapunov exponents vs.\ coupling strength $\sigma$ for the two-cluster CS pattern with $C_1=\{1,6\}$ and $C_2 = \{2,3,4,5,7,8,9,10\}$.
The exponent $\Lambda_{2}^{(1)}$ (red curve) is for cluster $C_1$, while $\Lambda_{2}^{(2)} = \Lambda_{3}^{(2)} = \Lambda_{4}^{(2)}$ (blue curve) and $\Lambda_{5}^{(2)} = \cdots = \Lambda_{8}^{(2)}$ (magenta curve) are for cluster $C_2$.
We use $\beta = \frac{2\pi}{3}-8\sigma$ and $\delta = \frac{\pi}{6}$.
Taking $\sigma = -0.37$ (downward arrow) makes $C_1$ unstable and $C_2$ stable.}
\label{Fig:EO_regular_N_10_k_8_N1_2_cluster_Lyapunov_exponents}
\end{figure}

\begin{figure}[h]
\includegraphics[width=0.85\linewidth]{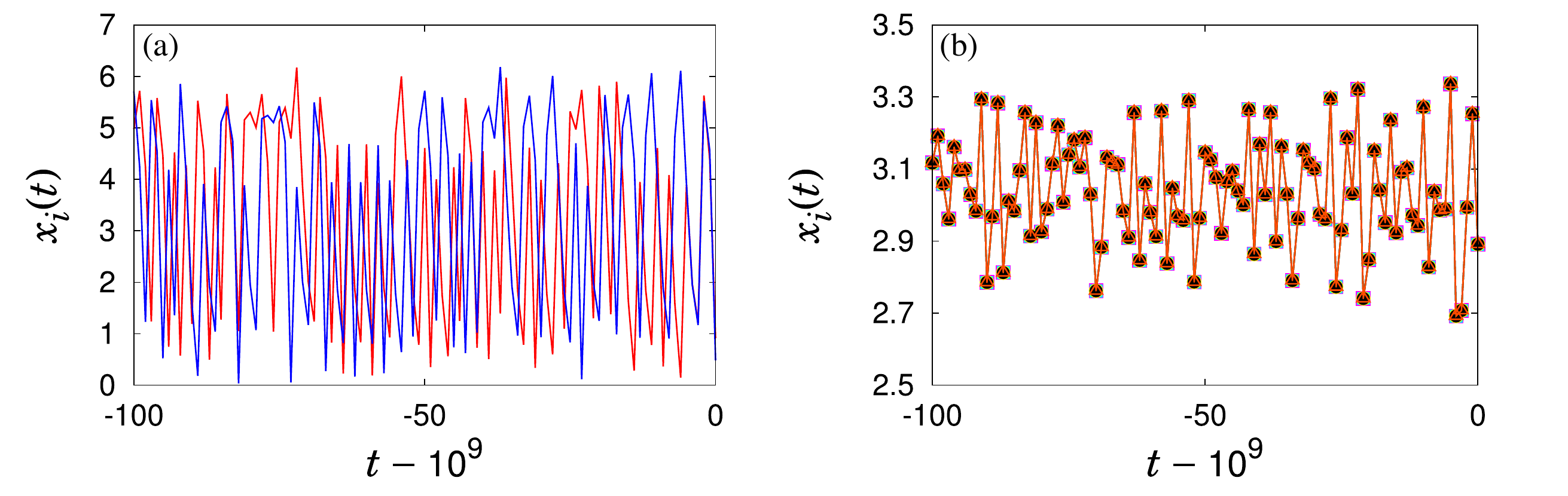}
\vspace{-8pt}
\caption{Chimera state of a ten-node ring network with a two-node incoherent cluster $C_1=\{1,6\}$ and an eight-node stably synchronized cluster $C_2=\{2,3,4,5,7,8,9,10\}$.
We plot $x_i(t)$ for the nodes in (a) $C_1$ and in (b) $C_2$, showing only the last $100$ iterations of a trajectory of length $10^9$.
We use parameters $\sigma = -0.37, \beta = \frac{2\pi}{3}-8\sigma$, $\delta = \frac{\pi}{6}$, and an initial condition with $x_i(0) \approx 1$ for $i \in C_1$ and $x_i(0) \approx 2$ for $i \in C_2$.}
\label{Fig:EO_regular_N_10_k_8_N1_2_dynamics}
\end{figure}

\begin{figure}[h]
\includegraphics[width=0.7\linewidth]{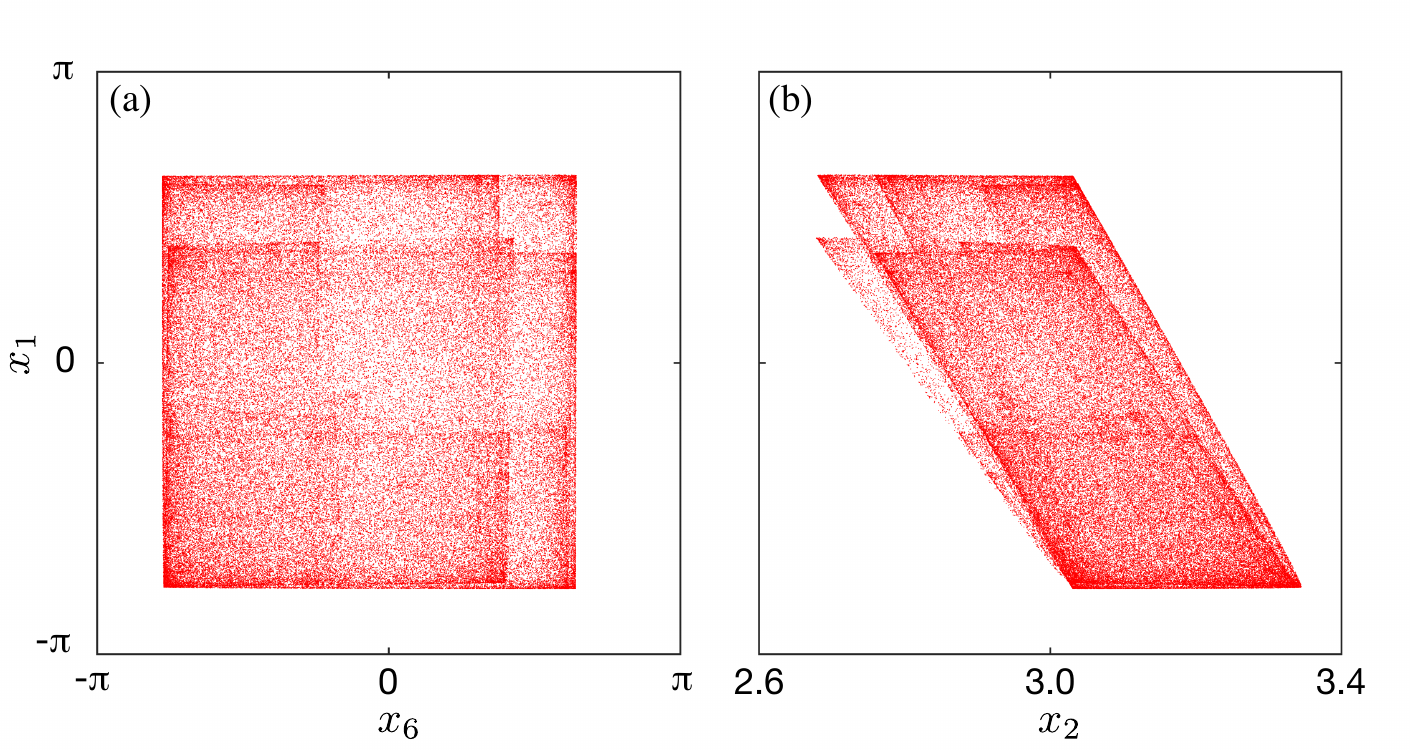}
\vspace{-8pt}
\caption{Chaotic dynamics of the chimera state in Fig.~\ref{Fig:EO_regular_N_10_k_8_N1_2_dynamics}.
(a) Projection onto the $(x_1,x_6)$-plane, showing the dynamics of nodes in $C_1$.
(b) Projection onto the $(x_1,x_2)$-plane, showing the dynamics a node in $C_1$ against a node in the other cluster, $C_2$.
Only the last $10^5$ points are shown in the trajectory of length $10^9$.}
\label{Fig:EO_regular_N_10_k_8_N1_2_chaotic}
\end{figure}

Here we consider the network with $N = 2k = 10$ nodes.
Setting $k=5$ in Eq.~\eqref{eqn:candidate_param}, we see that complete synchronization is unstable if $\beta = \frac{2\pi}{3} - 8\sigma$, $\delta = \frac{\pi}{6}$, and either $\sigma < -\frac{6-\pi}{15} \approx -0.19$ or $\sigma > \frac{6+\pi}{15} \approx 0.61$.
We consider the candidate two-cluster CS pattern with $k_1 = 1$ and $k_2 = 4$.
This pattern consists of two clusters, $C_1=\{1,6\}$ and $C_2 = \{2,3,4,5,7,8,9,10\}$, which are confirmed to be ISCs because the existence of the subgroups $G_1=\langle(1,6)\rangle$ and $G_2=\langle(2,3,4,5,7,8,9,10)\rangle$, respectively.
The corresponding synchronous state, $s_1(t)$ and $s_2(t)$, satisfy Eq.~\eqref{eqn:EO_reduced_two_cluster}.
The synchronization stability is determined by the transverse Lyapunov exponents $\Lambda_{2}^{(1)}$, $\Lambda_{2}^{(2)} = \Lambda_{3}^{(2)} = \Lambda_{4}^{(2)}$, and $\Lambda_{5}^{(2)} = \cdots = \Lambda_{8}^{(2)}$, which we compute numerically from Eq.~\eqref{eqn:EO_reduced_two_cluster} and Eqs.~(\ref{eqn:stabilty_2c_1}--\ref{eqn:stabilty_2c_4}).
The exponents are shown in Fig.~\ref{Fig:EO_regular_N_10_k_8_N1_2_cluster_Lyapunov_exponents} as functions of $\sigma$.
Taking $\sigma = -0.37$, we observe $\Lambda_{2}^{(1)}>0$, $\Lambda_{2}^{(2)}<0$, and $\Lambda_{5}^{(2)}<0$, which makes the synchronization of $C_1$ unstable while keeping the synchronization of $C_2$ stable.
We see that condition~(ii) for permanent chimera states is satisfied, since $C_1$ has only two nodes and cannot be further partitioned into nontrivial clusters.
Starting a trajectory close to the cluster synchronization manifold [i.e., $x_i(0) \approx 1$ for $i \in C_1$ and $x_i(0) \approx 2$ for $i \in C_2$], we indeed observe a chimera state that appears to be permanently stable, in which $C_1$ (i.e., $x_1$ and $x_6$) exhibit two-dimensional chaos and all nodes in $C_2$ are (chaotically) synchronized, as shown in Figs.~\ref{Fig:EO_regular_N_10_k_8_N1_2_dynamics} and \ref{Fig:EO_regular_N_10_k_8_N1_2_chaotic}.

\clearpage
\setlength{\textheight}{11in}

\begin{thebibliography}{99}
\bibitem{note2} Note that stable chimeras had been analytically established for infinite-size networks; see, e.g., Refs.~\cite{Abrams_chimera,PRL_Abraham}.
\bibitem{note1} The main difference is that the diagonalization of $A$ would be replaced by a Jordan canonical form in the general case of directed networks.
\end{thebibliography}
\end{document}